\documentclass[twocolumn]{aastex701}
\usepackage{hyperref}
\usepackage{amsmath}
\usepackage{subfigure}

\usepackage{graphicx,subcaption} 

\linenumbers

\begin{document}

\title{Probing Heavily Obscured AGN in Major Galaxy Mergers Using the mm–X-ray Correlation}

\author[0009-0007-7258-1261]{Macarena Droguett-Callejas}
\affiliation{Instituto de Astrof\'isica, Facultad de F\'isica, Pontificia Universidad Cat\'olica de Chile, Casilla 306, Santiago 22, Chile}
\email{mjdroguett@uc.cl}

\author[0000-0001-7568-6412]{Ezequiel Treister}
\affiliation{Instituto de Alta Investigaci{\'{o}}n, Universidad de Tarapac{\'{a}}, Casilla 7D, Arica, Chile}
\email{etreister@academicos.uta.cl}

\author[0000-0003-0057-8892]{Loreto Barcos-Mu\~noz}
\affiliation{National Radio Astronomy Observatory, 520 Edgemont Road, Charlottesville, VA 22903, USA}
\affiliation{Department of Astronomy, University of Virginia, 530 McCormick Road, Charlottesville, VA 22903, USA}
\email{lbarcos@nrao.edu}

\author[0000-0001-7690-3976]{Makoto Johnstone}
\affiliation{Department of Astronomy, University of Virginia, 530 McCormick Road, Charlottesville, VA 22903, USA}
\email{fhh3kp@virginia.edu }

\author[0000-0002-8686-8737]{Franz E. Bauer}
\affiliation{Instituto de Alta Investigaci{\'{o}}n, Universidad de Tarapac{\'{a}}, Casilla 7D, Arica, Chile}
\email{fbauer@academicos.uta.cl}

\author[0000-0002-6808-2052]{Taiki Kawamuro}
\affiliation{Department of Earth and Space Science, Osaka University, 1-1 Machikaneyama, Toyonaka 560-0043, Osaka, Japan}
\affiliation{RIKEN Cluster for Pioneering Research, 2-1 Hirosawa, Wako, Saitama, Saitama 351-0198, Japan}
\email{kawamuro@ess.sci.osaka-u.ac.jp}

\author[0000-0003-2638-1334]{N{\'u}ria Torres-Alb{\`a}}
\thanks{GECO Fellow}
\affiliation{Department of Astronomy, University of Virginia, 530 McCormick Road, Charlottesville, VA 22903, USA}
\affiliation{Department of Physics and Astronomy, Clemson University, Kinard Lab of Physics, Clemson, SC 29634-0978, USA}
\thanks{GECO Fellow}
\email{nuria@virginia.edu}

\author[0000-0001-5231-2645]{Claudio Ricci}
\affiliation{Department of Astronomy, University of Geneva, ch. d'Ecogia 16,1290, Versoix, Switzerland}
\affiliation{Instituto de Estudios Astrof\'isicos, Facultad de Ingenier\'ia y Ciencias, Universidad Diego Portales, Av. Ej\'ercito Libertador 441, Santiago, Chile}
\email{claudio.ricci.astro@gmail.com}

\author[0000-0002-7998-9581]{Michael Koss}
\affiliation{Eureka Scientific, 2452 Delmer Street, Suite 100, Oakland, CA 94602-3017, USA}
\affiliation{Space Science Institute, 4750 Walnut Street, Suite 205, Boulder, CO 80301, USA}
\email{Mike.Koss@eurekasci.com}

\author[0000-0002-3139-3041]{Yiqing Song}
\affiliation{European Southern Observatory, Alonso de Córdova 3107, Vitacura, Santiago Metropolitan, Chile.}
\email{Yiqing.Song@eso.org}

\author[0000-0003-2196-3298]{Alessandro Peca}
\affiliation{Eureka Scientific, 2452 Delmer Street, Suite 100, Oakland, CA 94602-3017, USA}
\affiliation{Department of Physics, Yale University, P.O. Box 208120, New Haven, CT 06520, USA}
\email{peca.alessandro@gmail.com}

\author[0000-0003-2638-1334]{Aaron Evans}
\affiliation{National Radio Astronomy Observatory, 520 Edgemont Road, Charlottesville, VA 22903, USA}
\affiliation{Department of Astronomy, University of Virginia, 530 McCormick Road, Charlottesville, VA 22903, USA}
\email{aevans@nrao.edu}

\author[0000-0003-3926-1411]{Jorge Gonz\'alez}
\affiliation{Instituto de Astrof\'isica, Facultad de F\'isica, Pontificia Universidad Cat\'olica de Chile, Casilla 306, Santiago 22, Chile}
\email{jgonzal@astro.puc.cl}




\begin{abstract}
The study of heavily obscured supermassive black hole (SMBH) growth in late-stage galaxy mergers is challenging: column densities $N_{\mathrm{H}}$$>$10$^{24}$cm$^{-2}$ can block most nuclear emission, leaving significant gaps in the SMBH growth census. Millimeter-wave continuum emission offers a potential window into this obscured phase, as it can trace Active Galactic Nuclei (AGN) activity through mechanisms less affected by dust extinction. In this work, we test whether the observed correlation between millimeter ($\sim$200~GHz) and hard X-ray (14–150~keV) luminosities can be used to plausibly identify hidden AGN in local (Ultra)Luminous Infrared Galaxies (U)LIRGs, including systems hosting confirmed dual AGN. We identify three sources — one confirmed AGN and two strong candidates — presenting significant evidence of AGN activity. The confirmed dual AGN lie within $\sim$3$\sigma$ of the mm–X-ray correlation, suggesting this relation can be used to identify hidden pairs. By combining the position of each source relative to this correlation with independent star formation rate constraints, we propose a method to disentangle AGN and star formation contributions for sources with measured column densities. While our analysis is based on a small, heterogeneous local sample and relies on empirical scaling relations, these results indicate that millimeter continuum emission may provide a useful complementary diagnostic for obscured SMBH growth. ALMA observations at high angular resolutions are particularly valuable for this approach, while future facilities such as the ngVLA will be essential to test its robustness in larger and more distant samples.

\end{abstract}



\section{Introduction} \label{sec:intro}
Major galaxy mergers can simultaneously trigger intense accretion onto supermassive black holes (SMBHs; e.g., \citealt{Hopkins_2008, Treister_2012}) while obscuring this activity behind large amounts of gas and dust, both on nuclear \citep[e.g.,][]{Ricci_2017, Koss_2018, Ricci_2021} and galaxy-wide \citep{Buchnner_2017, Blecha_2018} scales.  These critical events have also been linked to a high incidence of dual or closely separated AGN \citep[e.g.,][]{Imanishi2009, Pfeifle2019a, Pfeifle2019b, Farrah2022}. As a result, Active Galactic Nuclei (AGNs, i.e., rapidly accreting SMBHs) in late-stage mergers are often enshrouded in material that can reach Compton-thick ($N_{\mathrm{H}} > 10^{24}$ cm$^{-2}$; e.g., \citealt{Koss_2016, Yamada_2021, Ricci_2021}) levels, which hampers their detection and characterization. This obscuration creates a significant gap in the census of SMBH growth. A complete accounting of the dual AGN population is crucial for predicting gravitational wave event rates, enabling precise constraints on their number density and physical properties—measurements that will become feasible with next-generation space-based gravitational wave observatories \citep[e.g.,][]{Nandra_2013, Amaro-Seoane_2017, Piro_2023}.
Probing heavily obscured SMBH growth generally requires multi-wavelength observations \citep[e.g.,][]{Koss_2012}. In major galaxy mergers, however, intense star formation further complicates the AGN identification at optical, infrared, and radio wavelengths. For example, \citet{Koss_2011} reported a dual AGN detected in hard X-rays but missed in optical and radio observations, highlighting these difficulties. Optical emission-line diagnostics are also known to be unreliable in mergers due to extinction and dilution by star formation \citep{Koss_2010}.
Millimeter (mm) wavelengths offer a promising avenue to study AGN in such environments, as they are largely unaffected by dust, even at very high column densities ($N_{\mathrm{H}} \sim 10^{26}$ cm$^{-2}$; \citealt{Hildebrand_1983}). Detailed SED decomposition analyses \citep[e.g.,][]{Efstathiou2022, Varnava2025} have shown that many (U)LIRGs harbor deeply buried AGN, further motivating the use of dust-insensitive tracers such as mm emission. 

In most AGN, mm emission is thought to arise primarily from synchrotron radiation produced by relativistic particles near the X-ray–emitting corona \citep[e.g.,][]{Behar_2018, Jiang_2010}. Supporting this, \citet{Inoue_Doi_2018} detected excess mm emission in Seyfert galaxies, attributing it to self-absorbed synchrotron radiation from a compact region close to the SMBH. Because X-ray and mm emission can originate from related processes, they are expected to correlate tightly, a relation already explored in numerous studies \citep[e.g.,][]{Brinkmann_2000, Salvato_2004, Wang_2006, Panessa_2007, Behar_2018}.

More recently, \citet{Kawamuro_2022} investigated the relation between soft/ultra-hard X-ray (2–10 keV/14–150 keV) and mm-wave (1.3 mm, 230 GHz) luminosities for isolated AGN in the local Universe using the Swift/BAT 70-month catalog \citep{Baumgartner_2013, Ricci_2017b}. Swift/BAT’s sensitivity to ultra-hard X-rays (14–195 keV) ensures detections largely immune to obscuration up to Compton-thick levels ($N_{\mathrm{H}} \sim 10^{24}$ cm$^{-2}$), yielding one of the least biased local AGN samples. Using high angular resolution ($<0.6''$; $\lesssim$100–200 pc) ALMA observations, \citet{Kawamuro_2022} reported a tight correlation between mm and X-ray luminosities, with a 1$\sigma$ scatter of 0.36 dex. Because mm emission is optically thin, this correlation provides a powerful tool for studying SMBH growth in heavily obscured environments. Similarly, \citet{Ricci_2023} examined the 100 GHz (3.0 mm) to X-rays correlation for a smaller sample of nearby (within 50 Mpc) radio-quiet AGN from the BAT AGN Spectroscopic Survey (BASS; \citealt{Koss_2017}), finding even less scatter (0.22 dex) and no significant dependence of the intrinsic X-ray/mm ratio on AGN properties such as obscuration (in the Compton-thin regime), black hole mass, Eddington ratio, or star formation rate (SFR). However, its applicability beyond hard X-ray selected samples — in particular for relatively X-ray faint populations such as heavily reddened quasars \citep{Glikman_2024}, dust-rich Broad Absorption Line (BAL) systems \citep{Gallagher_2006}, e.g., LoBALs and FeLoBAL, infrared-luminous mergers, and/or extremely Compton-thick environments ($N_{\mathrm{H}}>10^{25}$ cm$^{-2}$) like the nuclear regions of gas-rich major mergers — remains largely unexplored.

(Ultra-)Luminous Infrared Galaxies or (U)LIRGs \citep{Sanders_1988}, provide an ideal laboratory to test this correlation. Defined by their infrared luminosities ($L_{\mathrm{IR}} > 10^{11} L_{\odot}$ for LIRGs and $L_{\mathrm{IR}} > 10^{12} L_{\odot}$ for ULIRGs), these sources often host galaxy mergers with both intense star formation and heavily-obscured AGN activity. Of particular interest is the dual AGN phase, in which late-stage mergers host two active nuclei separated by $<$10 kpc, frequently accompanied by extreme obscuration \citep{Ricci_2021} that can almost exclusively be probed at mm wavelengths.

In this work, we study 2–10 keV/14–150 keV X-ray and 1.3 mm luminosities for a sample of local (U)LIRGs from the Great Observatories All-Sky LIRG Survey (GOALS; \citealp{Armus_2009}). Some of these systems were selected as candidates to host obscured dual AGN, allowing us to test whether the mm–X-ray relation can reveal previously undetected AGN. The remaining sources are already confirmed dual AGN; to achieve a more complete sample of local confirmed dual AGN, we also include UGC 4211, a confirmed dual AGN not part of GOALS. Using these sources, we investigate archival Atacama Large Millimeter/submillimeter Array (ALMA) observations to examine whether the mm/X-ray relation holds in dual AGN systems and whether it can uncover hidden SMBH growth. Throughout this work we adopt a $\Lambda$CDM cosmology with $H_0 = 70$ km s$^{-1}$ Mpc$^{-1}$, $\Omega_{\mathrm{m}} = 0.3$, and $\Omega_{\Lambda} = 0.7$.

\section{Sample and Observational Data} \label{sec:Sample and Analysis}

The ultimate goal of this study is to evaluate the feasibility of using high angular resolution observations from ALMA, in combination with the millimeter–X-ray relation, to identify potential hidden AGN pairs in close proximity. This search for nearby obscured AGN using high angular resolution ALMA data will be presented in an upcoming publication (M. Droguett-Callejas et al., in prep.). As an initial step, and taking advantage of the currently available data, we focus on a sample of (U)LIRGs.
These sources are uniquely suited to probe the Compton-thick SMBH growth regimes and are prime environments for hosting heavily obscured (dual) AGN. We therefore searched the ALMA archive\footnote{\url{https://almascience.nrao.edu/aq/}} for observations of sources in the complete, volume-limited GOALS sample \citep{Armus_2009}, which contains more than 200 (U)LIRGs drawn from the IRAS Revised Bright Galaxy Sample \citep[RBGS;][]{Sanders_2003}. Specifically, we focused on observations from ALMA programs 2016.2.00055.S and 2017.1.00767.S (PI: E. Treister), designed to detect $^{12}$CO(2-1) emission and signatures of potential outflows in the line wings. These programs provide deep continuum data and a sample with well-defined statistical properties, selected based on the following criteria:
(1) $-20^\circ < \delta < 20^\circ$, to ensure good visibility with both ALMA and the VLA for potential follow-up; and 
(2) CO(1–0) peak flux $> 3$ mK, to guarantee detection of the $^{12}$CO(2–1) line . Note that these requirements imply that there will be (U)LIRGs included in the \citet{Kawamuro_2022} sample that will be automatically incorporated in this study, such as IRASF 05189-2524. The ALMA Band 6 (211–275 GHz) observations used in these projects provide an optimal balance of low optical depth, suitable spatial resolution, and data availability. To build our working sample, we further required sources with available X-ray detections in either the 2–10 keV or 14–150 keV bands. These criteria yield 11 systems observed with the Atacama Compact Array (ACA), with $\sim$5.4'' angular resolution, which correspond to a representative spatial resolution 
of $\sim$3.4 kpc for the sources in the sample. For four of them, CGCG 436$-$030, IRASF 14348$-$1447, IRASF 17138$-$1017, and IRASF 17207$-$0014, we also analyzed archival higher-resolution 12-m array data with a median angular resolution of $\sim$0.41'', corresponding to a typical spatial resolution of $\sim$0.33 kpc at the distance of these sources. While ACA observations have lower spatial resolution than the $<$1$''$ data used by \citet{Kawamuro_2022} for isolated AGN, they provide a valuable baseline for assessing how spatial resolution affects the detectability of obscured AGN.

Following \citet{Kawamuro_2022}, we searched the literature for X-ray data in both the 2–10 keV and 14–150 keV bands. The 2–10 keV data are primarily from {\it Chandra}/ACIS observations, whose relatively high spatial resolution reduces contamination from star-formation-related emission and can potentially resolve dual nuclei systems. For the 14–150 keV band, we used the Swift/BAT 105-Month Hard X-ray Survey \citep{Oh_2018}, since emission at these energies is considered a clean and relatively unbiased AGN tracer. Only one source in our sample, NGC 835, has a robust $L_{14–150,\mathrm{keV}}$ measurement, whereas IRASF 17138 has only an upper limit in the 14–195 keV energy range. Table \ref{tab:data_ULIRGS} summarizes the coordinates, and ALMA details for this (U)LIRG sample, including higher-resolution measurements where available.

\subsection{Dual AGN candidates}
he overarching objective of this paper is to assess the feasibility of applying the millimeter–X-ray relation to identify obscured AGN in general, and in particular in late-stage mergers. While this approach is applicable to a broad range of merger configurations and other obscured systems, we place particular emphasis on close-separation dual AGN systems, which represent a critical and observationally challenging subset for testing the power of this method. Hence, we compiled a sample that satisfies the following criteria: systems located in the local Universe ($z$$\leq$0.05), with projected nuclear separations between 60 pc and 10 kpc, available X-ray observations in both the 2–10 keV and 14–150 keV bands, and ALMA Band 6 (211–275 GHz) data obtained with the 12 meter array in extended configurations. The minimum nuclear separation was estimated using the maximum angular resolution achievable with ALMA Band 6 observations, combined with the mean distance of our sample and the minimum theoretical separation required to spatially resolve both nuclei.
To ensure completeness, we cross-checked our list with The Big Multi-AGN Catalog \citep[The Big MAC,][]{Pfeifle_2024}, the first literature-complete catalog of known confirmed and candidate multi-AGN systems. We hence verified that our subsample includes all currently confirmed dual AGN in the local Universe that satisfy our criteria, including UGC 4211, a confirmed dual AGN at very small nuclear separation not in GOALS. Our resulting sample comprises: Mrk 463, with a nuclear separation of $\sim 4$ kpc \citep{Treister_2018}; Mrk 739, with $\sim$3.4 kpc separation \citep{Koss_2011}; NGC 6240, with $\sim$1 kpc separation \citep{Treister_2020}; and UGC 4211, with $\sim$0.23 kpc separation \citep{Koss_2023}. For comparison, we also include the nearby (U)LIRG Arp 220, a late-stage merger hosting two nuclei separated by $\sim 0.36$ kpc \citep[e.g.,][]{Barcos-Munoz_2015}, where the presence of one or two AGN remains debated \citep[e.g.,][]{Soifer_1984, Becklin_1987, Iwasawa_2005, Barcos-Munoz_2015, Paggi_2017, Perna_2024}. Table \ref{tab:data_DUALS} lists these systems' properties and ALMA observational details.
We found X-ray data in the 2–10 keV range for all dual AGN in our sample, although UGC 4211 remains unresolved with {\it Chandra}. Four of the five systems also have measurements in the 14–150 keV band. Table \ref{tab:data_DUALS} summarizes the luminosity distances, coordinates, and ALMA data for our confirmed dual AGN/nuclei sample.

\begin{table*}[ht!]
\begin{centering}
\begin{tabular}{lcccccl} \hline 

 Source        &$z$ &
Spatial Scale [kpc/$''$]& R.A.\: [J2000]& DEC\: [J2000]&  Beam size &Project ID\\
 (1)&(2)&
(3)& (4)& (5)&  (6)&(7)\\  \hline \hline
 CGCG436-030&0.032 &0.66& 01:20:02.63&+14:21:42.3&   $6\farcs21 \times4\farcs 37$&2016.2.00055.S\\ 
 CGCG436-030 highres&''&''& ''& ''& $0\farcs 16 \times 0\farcs 13$ &$2017.1.01235.S^{*}$\\
 &  && & & &$2018.1.00279.S^{**}$\\   
 CGCG465-012&0.022 &0.46& 03:54:16.04& +15:55:43.4&  $6\farcs37 \times 4\farcs95$ &2016.2.00055.S\\   
 ESO550-IG025&0.032 &0.66& 04:21:20.04&  -18:48:45.2& $6\farcs 87 \times 3\farcs 74$ &2016.2.00055.S\\   
 IRAS18090+0130&0.029 &0.60& 18:11:33.34& +01:31:42.6&  $6\farcs 60 \times 4\farcs 36$&2017.1.00767.S\\   
 IRASF03359+1523&0.035 &0.72& 03:38:46.95&+15:32:54.5&  $6 \farcs 36 \times 4 \farcs 67$ &2017.1.00767.S\\   
 IRASF14348-1447&0.082 &1.60& 14:37:38.49& -15:00:19.1&  $6 \farcs 63 \times 4 \farcs 06$ &2017.1.00767.S\\   
 IRASF14348-1447 highres&''&''& ''& ''& $0 \farcs 04 \times 0 \farcs 04$ &2019.1.00329.S\\   
 IRASF17138-1017&0.017 &0.36& 17:16:35.82& -10:20:40.2&  $7 \farcs 71 \times 3 \farcs 99$ &2017.1.00767.S\\   
 IRASF17138-1017 highres&''&''& ''& ''& $0 \farcs 23 \times 0 \farcs 22$&2017.1.00255.S\\   
 IRASF17207-0014&0.043 &0.88& 17:23:21.95& -00:17:00.7&  $6 \farcs 77 \times 4 \farcs 40$&2017.1.00767.S\\   
 IRASF17207-0014 highres&''&''& ''& ''& $0 \farcs 39 \times 0 \farcs 28$ &2018.1.01123.S\\   
 NGC835&0.013 &0.28& 02:09:24.61& -10:08:09.1& $5 \farcs 72 \times 3 \farcs 74$ &2016.2.00055.S\\  
 NGC835& '' & '' & '' & '' &  $1 \farcs 48 \times 1 \farcs 08$&2018.1.00657.S\\  
 NGC838&0.013 &0.28& 02:09:38.53& -10:08:48.1&  $6 \farcs 11 \times 3 \farcs 97$ &2016.2.00055.S\\   
 UGC02238&0.022 &0.46& 02:46:17.51& +13:05:44.6& $6 \farcs 99 \times 4 \farcs 44$  &2016.2.00055.S\\ \hline
\end{tabular}
\caption{Sample of low resolution (U)LIRGs from GOALS studied in this work: \textbf{(1)} Source Name, \textbf{(2)} Redshift, \textbf{(3)} Linear scale in kpc/$''$, \textbf{(4)} Right Ascension (J2000), \textbf{(5)} Declination (J2000), \textbf{(6)} Beam size in arcseconds, and \textbf{(7)} Project ID, $^{*}$ corresponds to data from Band 3 and $^{**}$ Band 7. The values for redshift \textbf{(3)}, RA \textbf{(4)} and DEC \textbf{(5)} were obtained from the \href{https://ned.ipac.caltech.edu}{NASA/IPAC Extragalactic Database (NED)}.}

\label{tab:data_ULIRGS}
\end{centering}
\end{table*}

\begin{table*}[ht!]
\begin{centering}
\begin{tabular}{lcccccl} \hline 

 Source        &$z$ &Spatial Scale [kpc/$''$] & R.A.\: [J2000]& DEC\: [J2000]&  Beam size &Project ID\\
 (1)&(2)&(3)& (4)& (5)&  (6) &(7)\\  \hline \hline
 Mrk463&0.051 &1.04& 13:56:02.87& +18:22:19.48& $0 \farcs 30 \times 0\farcs 17$ &2013.1.00525.S\\
 Mrk739& 0.031 &0.64& 11:36:29.33& 21:35:45.10& $0 \farcs 09 \times 0 \farcs 08$ &2023.1.01196.S\\   
 NGC6240&0.024 &0.50& 16:52:58.87& +02:24:03.33& $0 \farcs 04\times 0 \farcs 02$ &2015.1.00370.S\\ 
 Arp220&0.018 &0.38& 15:34:57.21& +23:30:13.26& $0 \farcs 03 \times 0 \farcs 02$ &2017.1.00042.S\\  
 UGC4211&0.036 &0.74& 08:04:46.38& +10:46:36.19& $0 \farcs 06 \times 0\farcs 07$   &2021.1.01019.S\\  \hline
\end{tabular}
\caption{Sample of confirmed dual nuclei mergers in the local Universe studied in this work: \textbf{(1)} Source Name, \textbf{(2)} Redshift, \textbf{(3)} Linear scale in kpc/$''$, \textbf{(4)} Right Ascension (J2000), \textbf{(5)} Declination (J2000), \textbf{(6)} Beam size in arcseconds, and \textbf{(7)} Project ID. The values for redshift \textbf{(3)}, RA \textbf{(4)} and DEC \textbf{(5)} were obtained from the \href{https://ned.ipac.caltech.edu}{NASA/IPAC Extragalactic Database (NED)}}.

\label{tab:data_DUALS}
\end{centering}
\end{table*}

\section{Analysis}

\subsection{Mm-wave Emission}
\label{analisis: mm-wave}

We used the Common Astronomy Software Applications (CASA) package \citep{CasaTeam_2022} for calibration, imaging, and analysis of the ALMA data. We obtained the data from the ALMA archive (see project codes in Tables \ref{tab:data_ULIRGS} and \ref{tab:data_DUALS}) and ran the {\it script\_for\_PI.py} pipeline script provided by the observatory to obtain calibrated measurement sets. We then created dirty cubes of the observed spectral windows to identify and remove line emission from the data. Subsequently, we generated continuum images with these line-free measurement sets, which were then corrected for primary beam efficiency. Since the emission from a potential AGN corona would be spatially unresolved by any available ALMA configurations, we analyze the primary beam-corrected images by measuring the peak flux density within an aperture around the peak source emission (or from each nucleus for the confirmed dual AGNs). By using the peak, instead of the integrated flux density, we aim to remove some potential contribution from more diffuse dust/SF emission. In the case of the ACA data, the large beam size means that this diffuse dust/SF emission is most likely contributing to (or potentially dominating) the observed mm emission.  Further analysis and discussion of this potential contribution are presented in section \ref{analisis: sfc}.

We measured the noise of the continuum maps in the non-primary-beam-corrected images by selecting a large aperture free of emission. To calculate the mm-wave luminosity peak, we applied the standard relation between flux density and luminosity, which requires the luminosity distance and observed frequency for each source. The uncertainties in luminosity were derived by propagating systematic errors in the flux density measurement, with the RMS noise added in quadrature. The flux calibration accuracy for each ALMA band is outlined in the ``ALMA Proposer's Guide'' \citep{Privon_2025}, which reports a 10$\%$ accuracy for Band 6.

This procedure was applied to all the data studied here; only the high angular resolution data of CGCG436-030 required an additional step. Since no {high angular resolution Band 6 observations are currently available, we estimated the corresponding peak flux density at these wavelengths by interpolating the peak emission from archival Band 3 and 7 observations (after matching their angular resolution). We used the derived spectral index to compute the corresponding interpolated flux density emission at Band 6. Finally, for those galaxies with available high angular resolution complementary data, we compute the recovered emission fraction relative to the compact array measurements. To achieve this, we will compare the peak luminosity at both resolutions. This will serve as an estimation of how much diffuse emission is found in these sources. The mm-wave luminosity values and their corresponding errors are provided in Tables \ref{tab:results_2} and \ref{tab:results_1}.

\subsection{X-ray Emission}\label{subsec: x-ray emission}
Following the work of \citet{Kawamuro_2022}, who investigated correlations in two X-ray energy ranges (2–10 keV and 14–150 keV), we searched for luminosity measurements in both bands for each source. As summarized in Tables~\ref{tab:results_2} and \ref{tab:results_1}, 2–10 keV luminosities are available for the entire sample, whereas only five sources are currently detected by \textit{Swift}/BAT in the 14–150 keV range. Since the 14–150 keV band provides a cleaner, less contaminated probe of the mm–X-ray correlation in AGN, we estimated high-energy luminosities for the remaining sources to assess their expected behavior at these energies. To estimate 14-150 keV luminosities, we used the 2–10 keV measurements and a photon index of $\Gamma = 1.74$, appropriate for star-formation–dominated X-ray emission and not corrected for any AGN contribution. This photon index is consistent with other reported values for high-mass X-ray binaries (HMXBs; e.g.,  \citealp{Done_2007, Persic_Rephaeli_2007, Mineo_2014, Seifina_2016, Sazonov_2017}). This value corresponds to the median photon index for GOALS galaxies with no indication of an AGN in any wavelength, as compiled from the analysis of \citet{Iwasawa_2011} and \citet{Torres_Alba_2018} thereby ensuring that the adopted correction reflects non-AGN X-ray emission, and could be taken as a conservative approach in our search for the presence of previously-hidden AGN, as will be described below. The photon index was obtained from fitting Chandra data of individual galaxies in the 2-7 keV range, where thermal emission is no longer dominant. For the median computation, all galaxies for which the photon index was not fit, but rather fixed (due to a low number of counts), were removed. The extrapolation to the 14–150 keV band was carried out using the HEASARC WebPIMMS tool\footnote{\url{https://heasarc.gsfc.nasa.gov/cgi-bin/Tools/w3pimms/w3pimms.pl}}. This estimation specifically implies: (1) For sources where the X-ray emission is dominated by star formation, the extrapolated luminosities should be considered as upper limits, since the presence of a high-energy cutoff, as generally observed in these systems \citep{zhang_1996, Coburn_2002,Fornasini_2022}, would result in lower luminosities in the 14-150 keV range. (2) For Compton-thin AGN, the extrapolation should provide reliable estimates. (3) For heavily-obscured, Compton-thick AGN, the extrapolated luminosities likely represent lower limits due to the presence of a (prominent) Compton reflection hump not included in our assumed spectrum \citep{Lightman_1988, George_1991, Boorman_2025}. Although the 14–150 keV band represents a cleaner regime for investigating the mm–X-ray correlation, our use of this band is based on a model-dependent extrapolation of the observed 2–10 keV luminosities (for the sources without 14-150 observations). Future and complementary observations with NuSTAR—which directly probes the 3–79 keV range—would enable a more robust, empirical assessment of this relation, particularly for heavily obscured sources. The resulting luminosities in both energy bands are presented in Tables~\ref{tab:results_2} and \ref{tab:results_1}.

\subsection{Star Formation Contribution}
\label{analisis: sfc}
To characterize the millimeter and X-ray emission in major galaxy mergers, it is essential to consider that these events are typically associated with intense starburst activity \citep{Sanders_1988, Sanders_1996, Lonsdale_2006, Hopkins_2008}. High angular resolution studies \citep[e.g.,][]{Kawamuro_2022, Ricci_2023} can isolate the AGN contribution, but at lower resolution, both AGN and star formation contribute to the observed mm and X-ray emission. It is, hence, essential to quantify the host galaxy’s star-formation-related emission when interpreting the observed luminosities. This is particularly relevant for the low-resolution observations of the (U)LIRGs included in our sample. 

A potential caveat in this estimation arises from the possibility of having intrinsically X-ray weak AGN emission in merging systems. While in this case these sources could, in principle, remain undetected by X-ray–based diagnostics, the fraction of X-ray-weak AGN, such as BAL AGN in (U)LIRGs, appears to be small \citep[e.g.,][]{Teng_2015}, thus suggesting that they are unlikely to represent a significant source of incompleteness in our sample. 

\subsubsection{Star Formation Contribution to the X-ray emission}
Several studies have investigated the relation between star formation rate and X-ray luminosity across different samples and energy ranges. For example, \citet{Persic_Rephaeli_2007} studied local star-forming galaxies and (U)LIRGs using IRAS and XMM data, deriving the 2–10 keV SFR–$L_X$ relation. \citet{Lehmer_2010} constrained the SFR–$L_X$ relation in the 2–8 keV band for LIRGs from the IRAS Revised Bright Galaxy Sample \citep[RBGS;][]{Sanders_2003}. \citet{Mineo_2014} extended these studies with a combined sample of late-type galaxies, (U)LIRGs, and star-forming galaxies from the Chandra Deep Fields (CDFs), finding a linear relation: $L_{0.5-8 keV} \approx A \times 10^{39} \cdot \text{SFR} (M_{\sun} \: yr^{-1})$, with $A \approx (4 \pm 0.4)$. They also re-derived $A$ for the results of \citet{Persic_Rephaeli_2007} and \citet{Lehmer_2010}, finding good overall agreement.

\citet{Ricci_2021} examined this same connection in their hard X-ray study of late-stage AGN mergers, using NuSTAR observations of (U)LIRGs from the GOALS sample. They compared the observed X-ray luminosities with those predicted from star formation alone, based on the SFR–X-ray relations of \citet{Ranalli_2003} and \citet{Lehmer_2010}. Their results show that systems without clear AGN signatures have X-ray luminosities that are consistent with, or even below, the values expected from star formation. This agrees with the findings of \citet{Torres_Alba_2018}, who reported a flattening of the SFR–X-ray relation in (U)LIRGs, likely due to increased obscuration within their star-forming regions. In contrast, sources that host an AGN systematically display X-ray luminosities in excess of those predicted by star formation alone.

While this relation applies to the 2–10 keV band, we would also like to assess the expected star formation contribution at higher energies. \citet{Persic_Rephaeli_2007} showed that in powerful (U)LIRGs, the 2–10 keV emission is dominated by young HMXBs. Assuming that HMXBs dominate the SFR-driven X-ray output, we adopt the same photon index of $\Gamma = 1.74$ (See Section \ref{subsec: x-ray emission}). Figure 1 presents the observed mm and X-ray values of our (U)LIRGs sample against their derived SFR, compared to different SFR-mm/X-ray correlations. We derived the SFR values for our sample of (U)LIRGs by following the expression given by \citet{Kennicutt_1998}, corrected for a \citet{Kroupa_2001} Initial Mass Function (IMF) where $SFR = 3.15 \times 10^{-44} L_{IR}$. We derived the integrated IR luminosities [8-1000 \textmu m] using the observed IR flux densities reported by \citet{Sanders_2003} and the equations reported in Table 1 from \citet{Sanders_1996}. Panels a) and b) show the observed X-ray values for our sample of (U)LIRGs in the 2-10 and 14-150 keV bands, respectively, against their derived SFR. From both of these panels, we can see that the \citet{Lehmer_2010} relation predicts a higher contribution to the X-ray emission from star formation. In addition, our results are consistent with the study by \citet{Ricci_2021}, which found that sources without a confirmed AGN also exhibit consistent or lower luminosities compared to the relation by \citet{Lehmer_2010}. Consistently, one of the sources confirmed to host an AGN by other AGN tracers, NGC 835, is found significantly above the relation.\\ 

\subsubsection{Star Formation Contribution to the mm emission} At millimeter wavelengths, \citet{Murphy_2012} found a correlation between SFR and millimeter emission at 33 GHz using a sample of extranuclear and nuclear star formation regions in nearby galaxies. This relation uses SFRs computed using a \citet{Kroupa_2001} IMF and considers both a free-free emission and a synchrotron emission component (see equation 10 in \citet{Murphy_2012}). We use the explicit frequency dependence of this correlation to extrapolate from 33 GHz, the frequency at which it was derived, to $\sim$230 GHz, the representative frequency of the ALMA band 6 observations. This is done considering a non-thermal radio index of $\alpha^{NT}$ = 0.85, based on the average spectral index found for star-forming regions in NGC 6946. \citet{Yun_2002} also described an SFR-mm correlation based on a sample of IR-selected dusty starburst galaxies. This correlation used a combination of free-free, synchrotron, and dust emission to model the mm emission, and the $L_{IR}$ to derive SFR using \citet{Kennicutt_1998} corrected by a \citet{Kroupa_2001} IMF. The entire radio-to-FIR continuum flux spectrum is given by:\\
\begin{align*}
& S_{\nu_{obs}} =  \left\{ 25 f_{nth} \nu_{0}^{-\alpha} + 0.71 \nu_  {0}^{-0.1} \right. \\ 
& \left. + 1.3 \times 10^{-6} \frac{\nu_{0}^{3}\left [ 1 - e^{-(\nu_{0}/2000)^{1.35}} \right ]}{e^{0.00083\nu_{0}} - 1}\right\} \frac{(1+z) SFR}{D_{L}^{2}} Jy
\end{align*}
This radio-to-FIR continuum spectrum is modeled with $f_{\mathrm{nth}} = 1$ (Galactic normalization; \citealt{Condon_1992}) and a synchrotron spectral index of $\alpha = 0.75$ \citep{Yun_2002}. Emission is modeled for a range of SFRs using the GOALS sample's median redshift and luminosity distance, adopting $\nu_{0} = 230,\mathrm{GHz}$ as the representative ALMA Band 6 frequency. \\

Panel (c) in Figure \ref{fig:1} shows the location of our (U)LIRGs sample in the SFR-mm plane, compared with the two expected correlations described above. We can see here that the (U)LIRGs in our sample are located systematically below the relation derived by \citet{Yun_2002} and are instead better characterized by the correlation presented by \citet{Murphy_2012}. The fundamental difference between these two correlations is that the latter does not account for the dust contribution, whereas the former was derived from observations of dusty starburst galaxies at high redshifts.

\section{Results and Discussion} \label{Results and Discussion}

\begin{figure*}
    \centering 
    \subfigure[]{\includegraphics[width=0.4\textwidth]{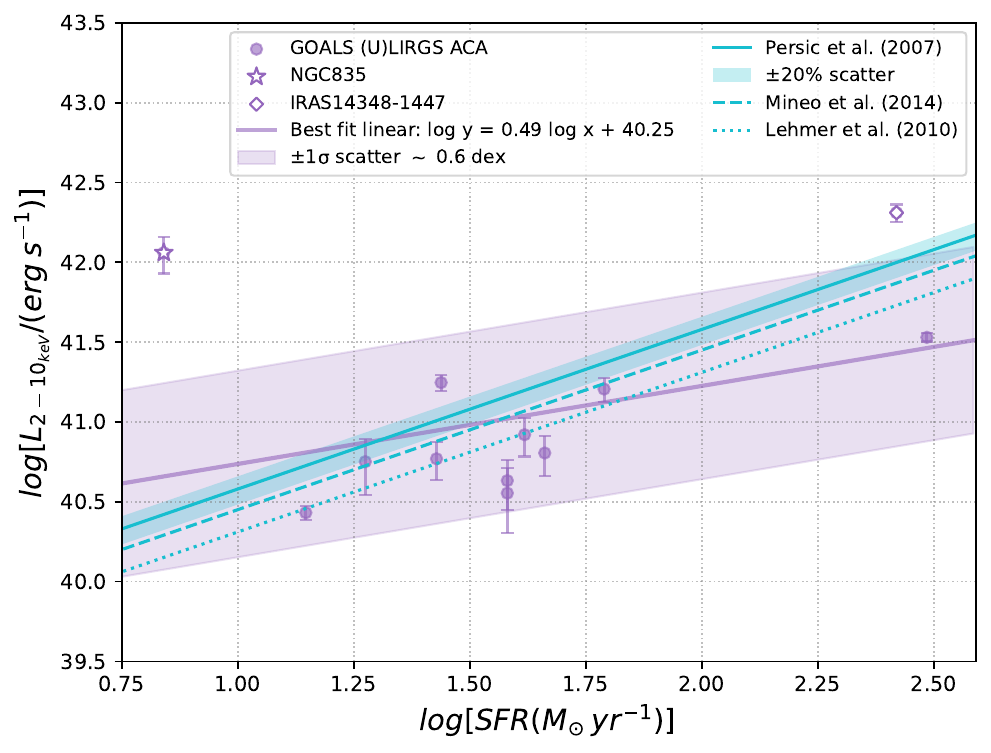}}
    \subfigure[]{\includegraphics[width=0.4\textwidth]{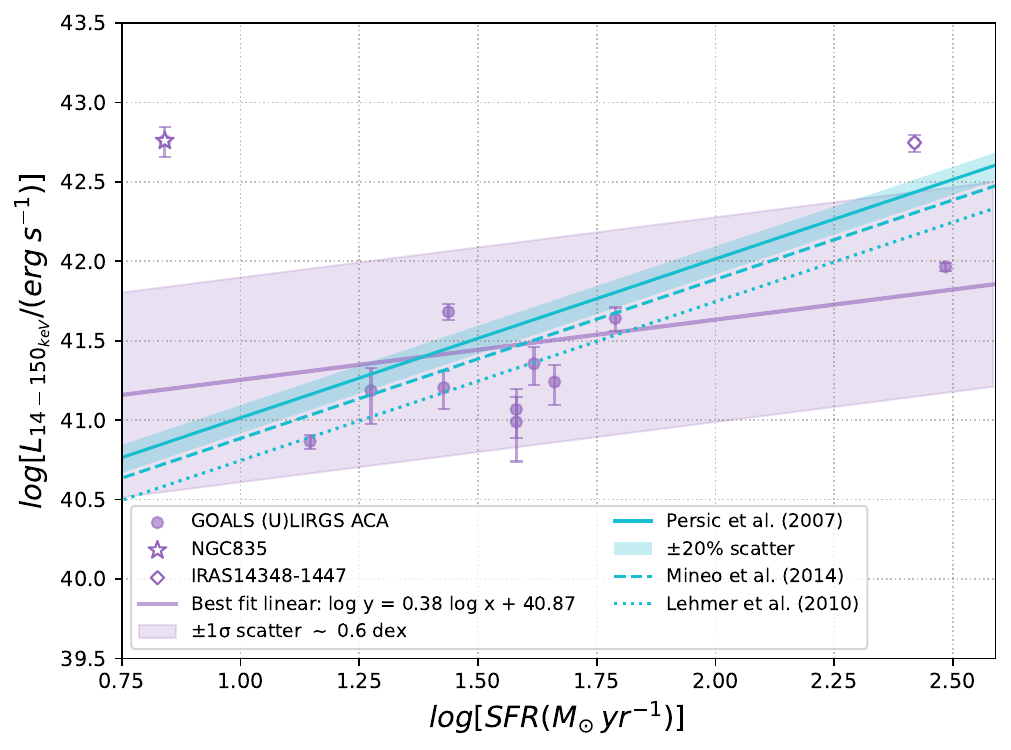}}\par
    \subfigure[]{\includegraphics[width=0.4\textwidth]{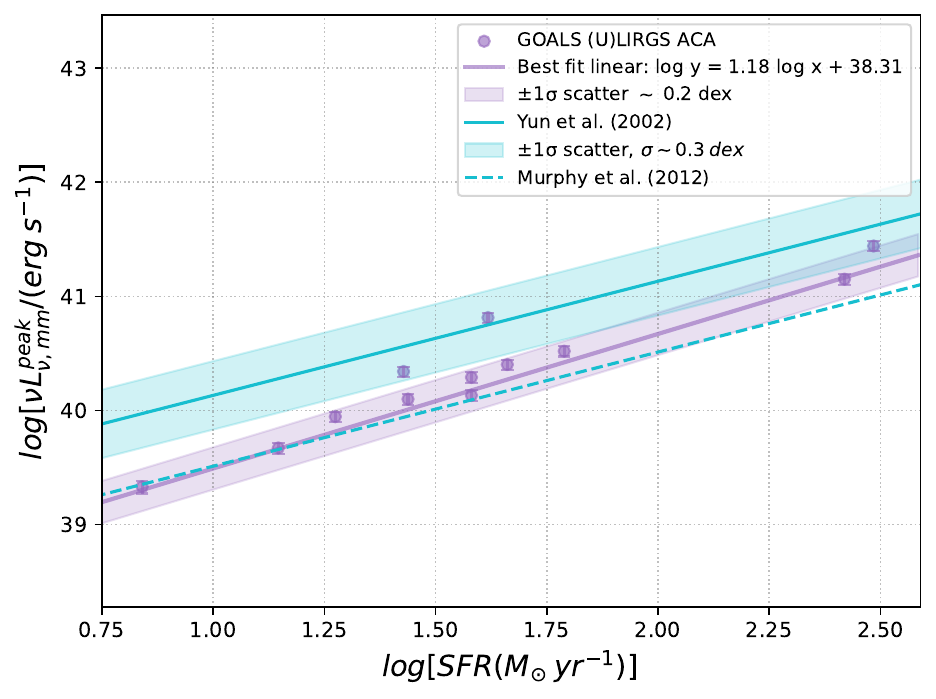}}
    \caption{Mm/x-ray luminosity versus SFR. Purple markers indicate the observed luminosity of the (U)LIRGs in our sample, plotted against their SFR derived from their IR luminosity. Light-blue lines show different SFR-mm/x-ray correlations. The grey dotted lines show a reference grid and are included solely to aid in the visual interpretation.  (a) This panel compares the position of the (U)LIRGs sources in the SFR-X-ray (2-10 keV) against the correlations derived by \citet{Persic_Rephaeli_2007}, derived from a sample of local star-forming galaxies,
    \citet{Mineo_2012}, obtained from a sample of 29 nearby late-type galaxies, (U)LIRGs and star forming galaxies, and \citet{Lehmer_2010}, measured from a sample of LIRGs at $z$$\sim$0. (b) Same as a) but for X-rays in the 14-150 keV range. (c) This panel shows the (U)LIRGs from our sample in the SFR-mm plane against the correlation derived by \citet{Yun_2002} derived from a sample of IR-selected dusty starburst galaxies and the correlation derived by \citet{Murphy_2012} , obtained from observations of nuclear and extra-nuclear star-forming regions in nearby galaxies.}
    \label{fig:foobar}
    \label{fig:1}
\end{figure*}

\begin{figure*}
    \centering
    \includegraphics[width=0.7\linewidth]{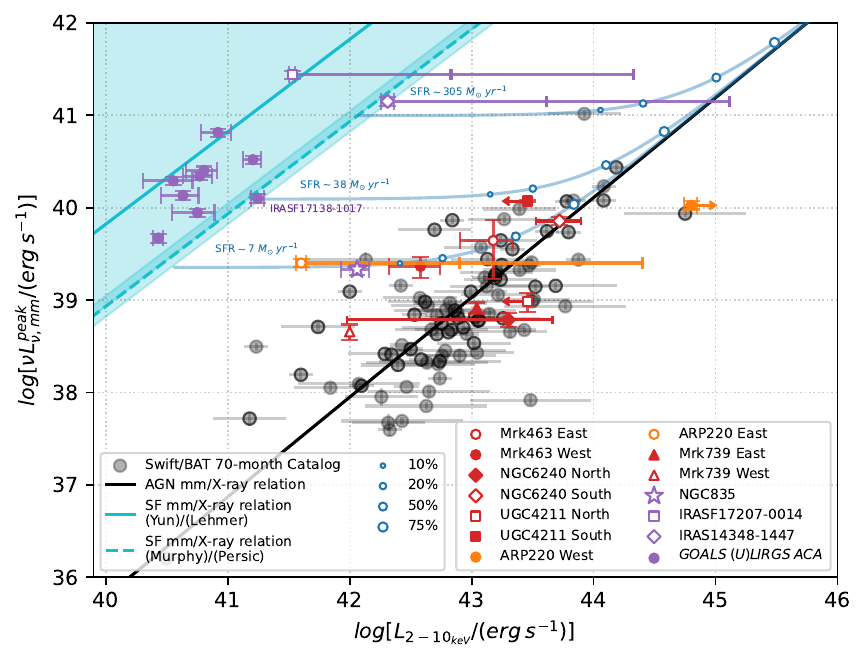}
    \includegraphics[width=0.7\linewidth]{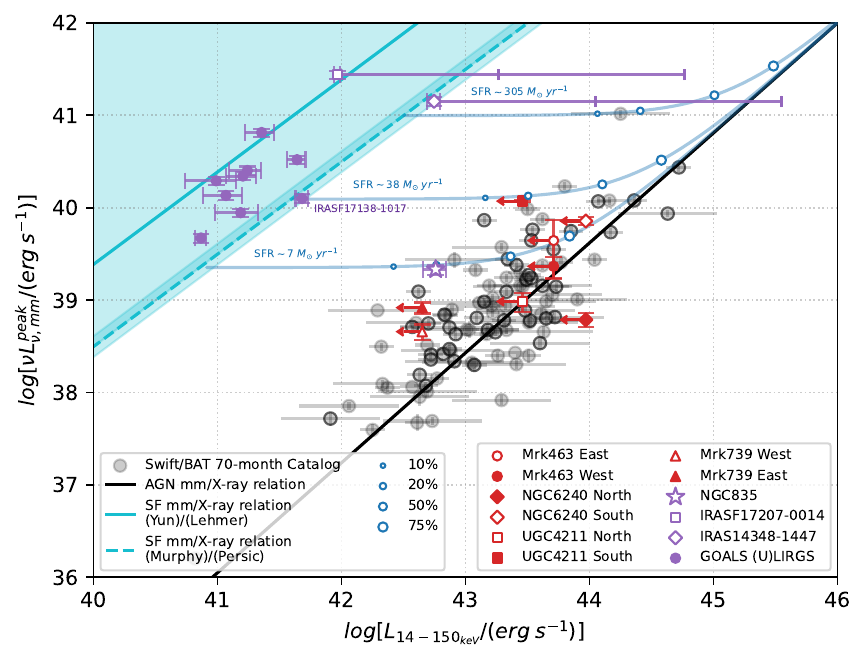}
    \caption{Observed peak millimeter continuum luminosity at 230 GHz versus X-ray luminosities in the 2–10 keV (top) and 14–150 keV bands (bottom). As in Figure \ref{fig:1}, the grey dotted lines show a reference grid and are included solely to aid in the visual interpretation. {\it Grey circles} show the isolated AGN sample from \citet{Kawamuro_2022}, with their mm–X-ray relation plotted with a {\it black line}. Sources highlighted with black-edged markers correspond to single AGN in \citet{Kawamuro_2022} sample that also satisfy criterion (1) of our sample. {\it Purple circles} represent the GOALS-selected IR-bright galaxies observed at low resolution with ACA; confirmed dual AGNs are marked in {\it red}, and Arp 220 in {\it orange}. When available, X-ray luminosities are corrected for obscuration using the reported column densities (Tables~\ref{tab:results_1} and \ref{tab:results_2}). For sources with $N_{\mathrm{H}}$ given as a range (Arp 220 E, IRASF 14348$-$1447, IRASF 17207$-$0014), a line of the same color indicates the shift in position if corrected for $N_{\mathrm{H}} \sim 10^{24}$ cm$^{-2}$ (first perpendicular bar) and $N_{\mathrm{H}} \sim 10^{25}$ cm$^{-2}$ (second perpendicular bar). The {\it light-blue shaded region} marks the area where emission can be entirely attributed to star formation, based on the SFR–mm relation from \citet{Yun_2002} and the SFR–X-ray relation from \citet{Persic_Rephaeli_2007}, as described in Section \ref{analisis: sfc}. {\it Light-blue lines} indicate the expected mm and X-ray luminosities for systems with different AGN contributions at a given SFR.}
     \label{fig:2}
\end{figure*}

\begin{table*}[ht!]
\centering
\begin{tabular}{lccclccc}
\hline
 Source        &$L_{mm}$   &RMS&Unresolved &$L_{2-10 keV}$& $L_{14-150 \: keV}$&& \\
  &$10^{40} \: erg \: s^{-1}$   &mJy/beam&Emission $\%$ &$10^{40} \: erg \: s^{-1}$&$10^{40} \: erg \: s^{-1}$ & $ \log N_{H}$&Reference  \\\hline \hline
 CGCG436-030&$3.31 \pm 0.34$&$0.11$&$\sim$ 22&$ 16.97 \pm 3.31$&$ 43.74 \pm 7.43$&-&a \\ 
 CGCG465-012&$0.88 \pm 0.09$&$0.11$&- &$ 5.64 \pm 0.88$& $ 15.35 \pm 5.86$&-&a \\ 
 ESO550-IG025  N&$1.95 \pm 0.20$&$0.14$&- &$ 3.58 \pm 1.95$&  $ 9.74 \pm 4.25$&-&a \\ 
 ESO550-IG025 S&$1.36 \pm 0.15$&$0.14$&- &$ 4.29 \pm 1.36$&  $ 11.68 \pm 3.99$&-&a \\ 
 IRAS18090+0130&$2.53 \pm 0.27$&$0.18$&- &$ 6.39 \pm 2.53$& $ 17.39 \pm 4.93$&-&a \\ 
 IRASF03359+1523&$6.52 \pm 0.66$&$0.17$&- &$8.32 \pm 6.52$&$ 22.65 \pm 6.10$&-&a \\ 
 IRASF14348-1447&$14.16 \pm  1.46$&$0.10$&$\sim$ 36&$ 204.66 \pm 25.62$& $ 557.08 \pm 69.72$&24/25&b \\ 
 IRASF17138-1017&$1.26 \pm 0.14$&$0.45$&$\sim$ 4&$ 17.6 \pm 1.26$& $ 48.02 \pm 5.51$&-&a \\ 
 IRASF17207-0014&$ 39.82 \pm  3.99 $&$0.30$&$\sim$ 19&$ 33.88 \pm 1.97$& $ 92.23 \pm 5.36$&24/25&b \\ 
 NGC835&$ 0.21 \pm  0.03 $&$0.24$&$\sim$ 32&$ 114.82 \pm 29.73$& $574.66_{-119.42}^{+123.20}$&$23.63_{-0.11}^{+0.13}$&b \\ 
 NGC838&$0.47 \pm 0.05$&$0.28$&- &$ 2.7 \pm 0.50$& $ 7.35 \pm 0.73$&-&a \\ 
 UGC02238&$2.20 \pm 0.22$&$0.11$&- &$ 5.88 \pm 2.20$& $ 16.01 \pm 4.29$&-&a \\ \hline
\end{tabular}
\caption{Derived parameters for the (U)LIRGs sources in our sample. The reference corresponds to the literature from where the values of the X-ray luminosity and $log_{10} \: N_{H}$; (a) corresponds to \cite{Torres_Alba_2018} and (b) to \cite{Ricci_2021}.}
\label{tab:results_2}
\end{table*}

\begin{table*}[ht!]
\centering
\begin{tabular}{lcclccc}
\hline
 Source        &$L_{mm}$  &RMS &$L_{2-10 keV}$& $L_{14-150 \: keV}$&&\\
&$ 10^{40} \: erg \: s^{-1}$  &mJy/beam&$10^{43} \: erg \: s^{-1}$&$ 10^{43}\: erg \: s^{-1}$ & $ \log N_{H}$& Reference \\\hline \hline
 Mrk463 E&$0.44^{+0.22}_{-0.41}$&$0.037$&$1.5 \pm 0.7$& $5.13^{+0.95}_{-2.00}$& $23.85 \pm 0.1$&a\\ 
 Mrk463 W&$0.23 \pm 0.06$&$0.037$&$0.38 \pm 0.17$& $5.13^{+0.95}_{-2.00}$&$23.51 \pm 0.1$&a\\
 Mrk739 W& $0.08 \pm 0.01$& $0.016$& $1.1 \pm 0.02$& $7.95^{+1.44}_{-0.99}$& $21.18^{+0.05}_{-0.06}$&b\\
 Mrk739 E& $0.05 \pm 0.01$& $0.016$& $0.1\pm 0.01$& $7.95^{+1.44}_{-0.99}$& $21.66^{+0.01}_{-0.01}$&b\\ 
  NGC6240 N &$0.06 \pm 0.01$&$0.029$&$2.00^{+0.36}_{-1.32} $& $9.33^{+0.14}_{-0.72} $&$24.19^{+0.17}_{-0.1}$&c\\
  NGC6240 S&$0.71 \pm 0.07$&$0.029$&$5.25^{+0.17}_{-0.20} $& $9.33^{+0.14}_{-0.72} $&$24.17^{+0.17}_{-0.32}$ &c\\
  Arp220 E&$0.25 \pm 0.03$&$0.020$&$>0.04$& -&-&d\\
  Arp220 W&$1.06 \pm 0.11$&$0.020$&$>0.1$& -& $>24.72 ^{\bigstar}$&d\\
  UGC4211 N&$0.09 \pm 0.02$&2&$0.5 \pm 1.6$& $2.88^{+1.11}_{-0.91}$&22.95$\pm$0.05$^{\ddagger}$&e\\
   UGC4211 S&$1.18 \pm 0.03$&15&$0.5 \pm 1.6$& $2.88^{+1.11}_{-0.91}$&22.95$\pm$0.05$^{\ddagger}$&e\\\hline
\end{tabular}

\caption{Derived parameters for the known dual nuclei or dual AGN in our sample. The references correspond to the existing literature from which the values of the X-ray luminosity and $log_{10} \: N_{H}$ were collected where (a) corresponds to \citet{Bianchi_2008}, (b) to \citet{Koss_2011},(c) to \citet{Ricci_2021}, (d) to \citet{Paggi_2017} and (e) to \citet{Koss_2023}. $^{\bigstar}$ value collected from \cite{Ricci_2021}. The value labeled with $^{\ddagger}$ was obtained from the work of \citet{Zhao_2021}.} 
\label{tab:results_1}
\end{table*}

Figure~\ref{fig:2} compares the X-ray and millimeter properties of our sample in order to assess the relative contributions of star formation and AGN activity. The top panel shows the observed 2–10 keV luminosities, while the bottom one presents the corresponding 14–150 keV luminosities, enabling a direct comparison between the empirical AGN and star-formation correlations.

The figure includes the empirical correlation and data for isolated AGN from \citet{Kawamuro_2022} in black. Some sources are highlighted with black-edged markers; these correspond to the single AGN in \citet{Kawamuro_2022} sample that also satisfy criterion (1) for our sample. We present a zone where emission can be explained {completely by star formation  (cyan, top left corner triangle). This zone is delimited by the \citet{Murphy_2012} and the \citet{Persic_Rephaeli_2007} relations. This combination was selected as a conservative approach, since it corresponds to the highest contribution from star formation to X-ray emission. We also show the relation of \citet{Yun_2002} as a reference for the expected contribution to the mm emission from dusty star-forming galaxies without AGN. We then expect sources in between the cyan region and the AGN mm-X-ray relation to have combine contributions from both star formation and a -possibly heavily obscured- AGN. To better understand cases with combined contributions, we have added light-blue lines, assuming a constant SFR and varying AGN contributions. For reference, we have marked the AGN contribution, as percentages (10, 20, 50, 75\%) of the total bolometric luminosity, with blue dots. Higher star formation rates produce bright mm emission, equivalent to that from bright, unobscured AGN. Higher star formation will hence naturally require more luminous AGNs to deviate from the correlation. In the vicinity of the AGN mm–X-ray correlation (black line), we expect AGN emission to dominate in both bands.\\


\subsection{(U)LIRGs}

In the top panel of Figure \ref{fig:2}, we present the observed 2–10 keV luminosities for all (U)LIRGs in our sample, with the exception of NGC~835, for which intrinsic (absorption-corrected) 2–10 keV luminosities are shown. The bottom panel shows the extrapolated 14–150 keV luminosities for the low-resolution (U)LIRGs (purple points). For systems lacking direct 14–150 keV measurements, the extrapolated luminosities should be regarded as upper limits of the star-formation contribution, since the high-energy cutoff, characteristic of star-formation-dominated emission, would in practice dramatically reduce the actual hard-X-ray emission (see Section~\ref{analisis: sfc}).

Most of our (U)LIRGs are found between the limits defined by the \citet{Yun_2002}/\citet{Lehmer_2010} and \citet{Murphy_2012} /\citet{Persic_Rephaeli_2007} correlations. These results suggest that most of these sources do not host a (dominating) AGN. This conclusion is based on measurements obtained at low spatial resolution for the majority of the GOALS (U)LIRGs in our sample. We caution, however, that since only observed X-ray luminosities are currently available for these sources (except for} IRASF17207-0014, IRASF14348-1447 and NGC835), we cannot rule out the presence of heavily obscured AGN that contribute to the mm continuum but remain faint in X-rays. Based on previously published results, the emission from most sources in this region is classified as Composite (AGN/SB), with no clear evidence for a dominant AGN. This is not surprising, as our method cannot easily distinguish the presence of a subdominant AGN either. ESO550-IG025 S is the only source in this region with an existing AGN claim \citep{Vardoulaki_2015}. Further verification based on other AGN tracers would be important to confirm the source of the emission in this system. 

An alternative explanation for the low X-ray luminosities of these systems is that they may be intrinsically X-ray faint, a phenomenon previously reported in (U)LIRGs \citep{Franceschini_2003, Lonsdale_2006, Ricci_2016}. Interestingly, a similar X-ray weakness is commonly observed in LoBAL quasars \citep{Green_1996}, which have been linked to (U)LIRGs, suggesting a possible evolutionary or physical connection between these populations. Still, the lack of additional AGN diagnostics argues against this scenario. For IRASF 17207-0014, the line-of-sight column densities from \citet{Ricci_2021} allow us to estimate possible intrinsic X-ray luminosities corrected for obscuration at $N_{\mathrm{H}} = 10^{24}$ cm$^{-2}$ and $10^{25}$ cm$^{-2}$, as presented in Table~\ref{tab:results_1}. In this case, the source could fall into the regime where the source has either a combination of star formation and AGN contribution, or it is mostly dominated by an obscured AGN (column densities of $N_{H} \sim 10^{25}$~cm$^{-2}$). However, no additional AGN evidence has been reported in the literature, and all available diagnostics indicate instead a starburst classification \citep{Iwasawa_2011}. A few (U)LIRGs lie in the intermediate region, with X-ray luminosities exceeding those expected from star formation alone. We interpret this excess as likely due to AGN activity. Indeed, NGC 835 and IRASF 14348-1447 are confirmed AGN hosts based on multiple diagnostics (see \citet{Vardoulaki_2015}, \citet{Song_2022} and references therein), while IRASF 17138-1017 shows evidence of an AGN only from its \textit{Chandra} hardness ratio \citep{Torres_Alba_2018}. \\ 

\subsection{Dual AGNs:} Confirmed dual AGNs are also presented in Figure \ref{fig:2}, using one symbol per nucleus, based on their inferred intrinsic X-ray luminosities. In UGC~4211, the 2–10 keV data cannot resolve the two nuclei, so the total intrinsic luminosity is assigned to each component and hence should be considered as an upper limit. Similarly, in the intrinsic 14–150 keV band, the total luminosity is treated as an upper limit for both components. Generally, the dual AGN systems are consistent with the mm–X-ray correlation within $\sim 3\sigma$ in both energy ranges. This is a promising result, as it confirms that the mm-X-ray relation can be effectively used to identify (obscured) growing SMBHs in (late-stage) major galaxy mergers. Finally, the dual nuclei in Arp 220 are shown as orange circles in the top panel. For the western nucleus, only a lower limit on the column density is available, and thus a lower bound is plotted for its X-ray luminosity. The eastern nucleus only has uncertain column density estimates, though previous studies \citep[e.g.,][]{Konig_2016} suggest that both nuclei would need to be heavily obscured to harbor AGN. The orange line indicates the expected conversion from observed to intrinsic luminosities for column densities between $10^{24}$ and $10^{25}$ cm$^{-2}$. We do not plot Arp 220 in the 14–150 keV panel, since the extrapolation would require assuming a spectral index that is not reliably constrained for such a heavily obscured system.

Overall, Figure \ref{fig:2} exhibits our approach: rather than correcting each individual source by subtracting the expected star formation contribution based on these relations, we instead define a cyan region (left-top light blue region in Figure \ref{fig:2}) that can be fully explained by star formation alone. Considering a region for star formation as opposed to correcting each source naturally accounts for uncertainties, such as those associated with observations and model predictions.

\subsection{Obscuration effects}
As discussed above, the mm continuum emission is expected to remain optically thin even in heavily obscured systems, whereas obscuration effects are more prominent in the X-ray regime. When column density measurements are available, we can correct the X-ray luminosities for obscuration. However, when $N_{H}$ is unavailable, we cannot estimate the true X-ray luminosity. Eight of the eleven GOALS sources lack column density estimates, which is expected given that in the low-counts regime, it is not feasible to carry out complex X-ray fitting to determine the presence of an AGN component. Their X-ray luminosities could therefore be underestimated; however, in most cases, there is no evidence for an AGN, and the emission is consistent with an origin due to star formation, for which extinction corrections are expected to be modest. These sources generally lie in the location expected for pure star-formation emission. Concealing an AGN in these systems would require extreme obscurations, with $N_{\mathrm{H}} > 10^{25}$ cm$^{-2}$; values that are rare in the general AGN population \citep[e.g.,][]{Peca_2024}, and/or intrinsically under-luminous X-ray emission. Nevertheless, X-ray studies of nearby (U)LIRGs in GOALS \citep{Ricci_2021} suggest that a significant fraction of late-stage mergers could harbor heavily obscured, as-yet undetected AGN—underscoring the importance of calibrating an alternative mm AGN tracer. For the (U)LIRGs within the cyan triangle, showing a significant mm excess above the star-formation prediction, one could attempt to estimate the AGN luminosity required to account for this excess, if a reliable determination of the dominant star-formation component is feasible.

For the systems in the confirmed dual AGN sample, each nucleus was independently corrected for obscuration using either the exact column density or an upper limit, as reported in Table~\ref{tab:results_1}. Arp~220 is expected to be highly obscured \citep[e.g.,][]{Iwasawa_2001, Teng_2015}, so further constraints on the column densities of each nucleus are needed. For instance, if the line-of-sight $N_{\mathrm{H}}$ for the eastern nucleus is confirmed to lie between $10^{24}$ and $10^{25}$~cm$^{-2}$, the source would align with the \citet{Kawamuro_2022} relation, potentially confirming the presence of AGN activity. This highlights the importance of obtaining improved column-density measurements in future work, as they are essential for accurately determining the intrinsic luminosities of the sources under study.

\subsection{Spatially-extended Continuum mm Emission} \label{spatial resolution}

AGN coronal millimeter continuum emission originates from a very compact region \citep[$\sim10^{-4}$–$10^{-3}$~pc;][]{Behar_2018}, which remains unresolved at the scales probed by our work and the literature. In contrast, extended millimeter emission on larger scales is more likely associated with star formation \citep{Panessa_2019} or dusty components with spatial extents of the order of tens of parsecs. Unfortunately, even with igh angular resolution observations, we would not be able to resolve these dusty components. 
The (U)LIRG sample studied here has an average ALMA ACA resolution of $\sim 5.4''$. To evaluate the effect of this coarser resolution, we analyzed four GOALS sources in our sample —CGCG436-030, IRASF 14348-1447, IRASF 17138-1017, and IRASF 17207-0014— for which higher-resolution ALMA 12-m array data are available. This high angular-resolution data has a median angular resolution of $\sim$0.4'', on average corresponding to spatial scales of $\sim$0.3 kpc.

Comparing the peak luminosity at high and low resolutions to that from lower-resolution ACA data, we find that the former are consistently lower, indicating the presence of diffuse SF mm emission in the latter. Table~\ref{tab:results_2} lists the fraction of luminosity attributed to unresolved emission (from the high resolution data), with a maximum of $\sim$36\%, indicating the presence of considerable extended emission coming from star formation in these four sources. Therefore, for the ACA (U)LIRG observations in our sample, even if a heavily obscured AGN is present, it would not be expected to dominate the mm continuum. The precise determination of the emission's origin requires further follow-up information. Unlike the Swift/BAT AGN hosted by relatively normal star-forming galaxies, used to derive the mm–X-ray relation of \citet{Kawamuro_2022}, GOALS galaxies often host luminous, compact starbursts that will need to be taken into consideration, as they can still be responsible for a significant fraction of the observed star-formation–related emission in high angular resolution observations. Hence, multi-wavelength, high angular resolution SED decompositions are crucial to further confirm and provide a more detailed study identifying the origin and quantifying the contribution of nuclear millimeter emission \citep{DelPalacio_2025}. By combining multi-band ALMA data with other high angular resolution facilities, such as the VLA and VLBI, it becomes possible to derive spatially-resolved spectral indices, offering stronger constraints on the physical mechanisms powering the emission.

\section{Conclusions} 

Here, we investigated whether the relationship between the 230 GHz mm continuum and X-ray luminosity (2–10 keV and 14–150 keV) can be used as a tracer of hidden SMBH growth, with the ultimate goal of evaluating this approach for identifying obscured dual AGN at small separations. To this end, we analyzed a sample of (U)LIRGs from the GOALS survey — including both ACA observations and high angular resolution confirmed dual AGN — as well as the confirmed dual AGN UGC4211 and the baffling late-stage merger Arp~220.
We compared the locus of these systems relative to the established mm–X-ray correlation for isolated AGN and star-forming galaxies.
The (U)LIRGs follow the relation of \citet{Murphy_2012}, which accounts only for free-free and synchrotron emission, rather than the \citet{Yun_2002} relation that includes dust emission. This suggests that the dust contribution to the 230 GHz continuum in our sample is relatively small, with a scatter of only $\sim$0.2dex in the mm–SFR plane. Rather than applying per-object corrections for star formation, we adopted a conservative approach by defining a region in the X-ray–mm luminosity plane that can be fully explained by star formation alone. Our results show that the position of galaxy nuclei relative to this region provides important constraints on the dominant physical processes:

\begin{itemize}
\item \textit{Star-formation locus:} Sources in the region expected for pure star formation can be explained without invoking AGN activity, unless the SMBH growth emission is extremely Compton-thick and/or intrinsically X-ray under-luminous. 

\item \textit{AGN correlation}: Sources lying near the AGN mm–X-ray relation of \citet{Kawamuro_2022} are most likely AGN-dominated, since star formation alone cannot reproduce their luminosity ratios.
\item \textit{Intermediate regime}: Sources between the star-formation locus (cyan triangle in Figure~\ref{fig:2}) and the AGN correlation require a combination of star formation and obscured AGN emission to explain their luminosities. In our sample, four (U)LIRGs fall in this category: IRASF17207$-$0014 and IRASF14348$-$1447 (requiring $N_{\rm H} \sim 10^{24}-10^{25}$ cm$^{-2}$), IRASF17138$-$1017 ($N_{\rm H} > 10^{24}$ cm$^{-2}$), and NGC835 (already corrected for obscuration). Among these, IRASF17207$-$0014 is classified as a starburst in multiple diagnostics, while NGC835 and IRASF14348$-$1447 are confirmed AGN hosts established by multiple AGN diagnostics. Our approach confirms the AGN nature of NGC835 while providing strong evidence for AGN activity in IRASF14348$-$1447 and IRASF17138$-$1017, by combining our approach with the observed \textit{Chandra} hardness ratio.
\end{itemize}

Overall, this analysis of our sample of (U)LIRGs from GOALS, plus UGC 4211, strongly suggests the presence of heavily obscured AGN in three sources, two of which already have significant evidence in the literature. The confirmed dual AGN in our sample lie within $\sim$3$\sigma$ of the mm–X-ray correlation, demonstrating that the X-ray–mm continuum correlation extends to dual SMBH systems and can aid in identifying hidden pairs.

We note that, unlike the Swift/BAT AGN hosted by relatively normal star-forming galaxies used to derive the mm–X-ray relation of \citet{Kawamuro_2022}, GOALS galaxies often host luminous, compact starbursts. These can contribute significantly to the observed mm emission even at high angular resolution, and multi-wavelength SED decompositions are essential for robustly confirming the origin of nuclear mm emission.

These encouraging results suggest that the X-ray–mm continuum correlation can be a useful tool for uncovering previously hidden SMBH growth in merging galaxies. ALMA's high sensitivity and spatial resolution make it particularly well-suited for this application, while future facilities such as the ngVLA will extend this approach to fainter and more distant systems.


\begin{acknowledgments}
We thank the anonymous referee for a detailed and constructive review of our work, which, in our opinion, significantly improved this manuscript.
We gratefully acknowledge funding from
ANID BECAS/DOCTORADO NACIONAL 21210485 (MDC), CATA-BASAL - FB210003 (MDC, ET, FEB), Millennium Science Initiative - AIM23-0001 and ICN12\_009 (FEB), and FONDECYT Regular - 1241005 and 1250821 (ET, FEB). ET would like to thank the generous hospitality of the North American ALMA Science Center (NAASC) at NRAO during his sabbatical stay in 2022, where a significant fraction of this work was carried out. CR acknowledges support from SNSF Consolidator grant F01$-$13252, Fondecyt Regular grant 1230345, ANID BASAL project FB210003, and the China-Chile joint research fund. This paper makes use of the following ALMA data: ADS/JAO.ALMA$\#$2016.2.00055.S, 
ADS/JAO.ALMA$\#$2013.1.00525.S,\\ 
ADS/JAO.ALMA$\#$2015.1.00370.S,\\
ADS/JAO.ALMA$\#$2017.1.00042.S,\\
ADS/JAO.ALMA$\#$2017.1.00255.S,\\ 
ADS/JAO.ALMA$\#$2017.1.00767.S,\\ 
ADS/JAO.ALMA$\#$2017.1.01235.S,\\ 
ADS/JAO.ALMA$\#$2018.1.00279.S,\\
ADS/JAO.ALMA$\#$2018.1.01123.S,\\
ADS/JAO.ALMA$\#$2019.1.00329.S,\\
ADS/JAO.ALMA$\#$2021.1.01019.S.\\ 
ALMA is a partnership of ESO (representing its member states), NSF (USA) and NINS (Japan), together with NRC (Canada), NSTC and ASIAA (Taiwan), and KASI (Republic of Korea), in cooperation with the Republic of Chile. The Joint ALMA Observatory is operated by ESO, AUI/NRAO and NAOJ. The National Radio Astronomy Observatory and Green Bank Observatory are facilities of the U.S. National Science Foundation operated under cooperative agreement by Associated Universities, Inc. This research has made use of the NASA/IPAC Extragalactic Database (NED), which is operated by the Jet Propulsion Laboratory, California Institute of Technology, under contract with the National Aeronautics and Space Administration.

\facilities{ALMA}

\software{NumPy \citep{2020NumPy-Array}, Matplotlib \citep{Matplotlib}, CASA \citep{CasaTeam_2022}}

\end{acknowledgments}

\bibliography{sample631}{}

@ARTICLE{Gallagher_2006,
       author = {{Gallagher}, S.~C. and {Brandt}, W.~N. and {Chartas}, G. and {Priddey}, R. and {Garmire}, G.~P. and {Sambruna}, R.~M.},
        title = "{An Exploratory Chandra Survey of a Well-defined Sample of 35 Large Bright Quasar Survey Broad Absorption Line Quasars}",
      journal = {\apj},
         year = 2006,
        month = jun,
       volume = {644},
       number = {2},
        pages = {709-724},
          doi = {10.1086/503762},
archivePrefix = {arXiv},
       eprint = {astro-ph/0602550},
 primaryClass = {astro-ph},
       adsurl = {https://ui.adsabs.harvard.edu/abs/2006ApJ...644..709G}
}

@ARTICLE{Glikman_2024,
       author = {{Glikman}, Eilat and {LaMassa}, Stephanie and {Piconcelli}, Enrico and {Zappacosta}, Luca and {Lacy}, Mark},
        title = "{Accretion and obscuration in merger-dominated luminous red quasars}",
      journal = {\mnras},
         year = 2024,
        month = feb,
       volume = {528},
       number = {1},
        pages = {711-725},
          doi = {10.1093/mnras/stae042},
archivePrefix = {arXiv},
       eprint = {2401.02859},
 primaryClass = {astro-ph.GA},
       adsurl = {https://ui.adsabs.harvard.edu/abs/2024MNRAS.528..711G}
}

@ARTICLE{Panessa_2019,
       author = {{Panessa}, Francesca and {Baldi}, Ranieri Diego and {Laor}, Ari and {Padovani}, Paolo and {Behar}, Ehud and {McHardy}, Ian},
        title = "{The origin of radio emission from radio-quiet active galactic nuclei}",
      journal = {Nature Astronomy},
         year = 2019,
        month = apr,
       volume = {3},
        pages = {387-396},
          doi = {10.1038/s41550-019-0765-4},
archivePrefix = {arXiv},
       eprint = {1902.05917},
 primaryClass = {astro-ph.GA},
       adsurl = {https://ui.adsabs.harvard.edu/abs/2019NatAs...3..387P}
}

@ARTICLE{Peca_2024,
       author = {{Peca}, Alessandro and {Cappelluti}, Nico and {LaMassa}, Stephanie and {Urry}, C. Megan and {Moscetti}, Massimo and {Marchesi}, Stefano and {Sanders}, David and {Auge}, Connor and {Ghosh}, Aritra and {Ananna}, Tonima Tasnim and {Torres-Alb{\`a}}, N{\'u}ria and {Treister}, Ezequiel},
        title = "{Stripe 82-XL: The {\ensuremath{\sim}}54.8 deg$^{2}$ and {\ensuremath{\sim}}18.8 Ms Chandra and XMM-Newton Point-source Catalog and Number of Counts}",
      journal = {\apj},
         year = 2024,
        month = oct,
       volume = {974},
       number = {2},
          eid = {156},
        pages = {156},
          doi = {10.3847/1538-4357/ad6df4},
archivePrefix = {arXiv},
       eprint = {2407.09617},
 primaryClass = {astro-ph.GA},
       adsurl = {https://ui.adsabs.harvard.edu/abs/2024ApJ...974..156P}
}

@ARTICLE{Ricci_2023,
       author = {{Ricci}, Claudio and {Chang}, Chin-Shin and {Kawamuro}, Taiki and {Privon}, George C. and {Mushotzky}, Richard and {Trakhtenbrot}, Benny and {Laor}, Ari and {Koss}, Michael J. and {Smith}, Krista L. and {Gupta}, Kriti K. and {Dimopoulos}, Georgios and {Aalto}, Susanne and {Ros}, Eduardo},
        title = "{A Tight Correlation between Millimeter and X-Ray Emission in Accreting Massive Black Holes from <100 mas Resolution ALMA Observations}",
      journal = {\apjl},
         year = 2023,
        month = aug,
       volume = {952},
       number = {2},
          eid = {L28},
        pages = {L28},
          doi = {10.3847/2041-8213/acda27},
archivePrefix = {arXiv},
       eprint = {2306.04679},
 primaryClass = {astro-ph.HE},
       adsurl = {https://ui.adsabs.harvard.edu/abs/2023ApJ...952L..28R}
}

@ARTICLE{Buchnner_2017,
       author = {{Buchner}, Johannes and {Bauer}, Franz E.},
        title = "{Galaxy gas as obscurer - II. Separating the galaxy-scale and nuclear obscurers of active galactic nuclei}",
      journal = {\mnras},
         year = 2017,
        month = mar,
       volume = {465},
       number = {4},
        pages = {4348-4362},
          doi = {10.1093/mnras/stw2955},
archivePrefix = {arXiv},
       eprint = {1610.09380},
 primaryClass = {astro-ph.HE},
       adsurl = {https://ui.adsabs.harvard.edu/abs/2017MNRAS.465.4348B}
}

@ARTICLE{Treister_2012,
       author = {{Treister}, E. and {Schawinski}, K. and {Urry}, C.~M. and {Simmons}, B.~D.},
        title = "{Major Galaxy Mergers Only Trigger the Most Luminous Active Galactic Nuclei}",
      journal = {\apjl},
         year = 2012,
        month = oct,
       volume = {758},
       number = {2},
          eid = {L39},
        pages = {L39},
          doi = {10.1088/2041-8205/758/2/L39},
archivePrefix = {arXiv},
       eprint = {1209.5393},
 primaryClass = {astro-ph.CO},
       adsurl = {https://ui.adsabs.harvard.edu/abs/2012ApJ...758L..39T}
}

@ARTICLE{Kawamuro_2022,
       author = {{Kawamuro}, Taiki and {Ricci}, Claudio and {Imanishi}, Masatoshi and {Mushotzky}, Richard F. and {Izumi}, Takuma and {Ricci}, Federica and {Bauer}, Franz E. and {Koss}, Michael J. and {Trakhtenbrot}, Benny and {Ichikawa}, Kohei and {Rojas}, Alejandra F. and {Smith}, Krista Lynne and {Shimizu}, Taro and {Oh}, Kyuseok and {den Brok}, Jakob S. and {Baba}, Shunsuke and {Balokovi{\'c}}, Mislav and {Chang}, Chin-Shin and {Kakkad}, Darshan and {Pfeifle}, Ryan W. and {Privon}, George C. and {Temple}, Matthew J. and {Ueda}, Yoshihiro and {Harrison}, Fiona and {Powell}, Meredith C. and {Stern}, Daniel and {Urry}, Meg and {Sanders}, David B.},
        title = "{BASS XXXII: Studying the Nuclear Millimeter-wave Continuum Emission of AGNs with ALMA at Scales {\ensuremath{\lesssim}}100-200 pc}",
      journal = {\apj},
         year = 2022,
        month = oct,
       volume = {938},
       number = {1},
          eid = {87},
        pages = {87},
          doi = {10.3847/1538-4357/ac8794},
archivePrefix = {arXiv},
       eprint = {2208.03880},
 primaryClass = {astro-ph.GA},
       adsurl = {https://ui.adsabs.harvard.edu/abs/2022ApJ...938...87K}
}

@ARTICLE{Koss_2023,
       author = {{Koss}, Michael J. and {Treister}, Ezequiel and {Kakkad}, Darshan and {Casey-Clyde}, J. Andrew and {Kawamuro}, Taiki and {Williams}, Jonathan and {Foord}, Adi and {Trakhtenbrot}, Benny and {Bauer}, Franz E. and {Privon}, George C. and {Ricci}, Claudio and {Mushotzky}, Richard and {Barcos-Munoz}, Loreto and {Blecha}, Laura and {Connor}, Thomas and {Harrison}, Fiona and {Liu}, Tingting and {Magno}, Macon and {Mingarelli}, Chiara M.~F. and {Muller-Sanchez}, Francisco and {Oh}, Kyuseok and {Shimizu}, T. Taro and {Smith}, Krista Lynne and {Stern}, Daniel and {Tello}, Miguel Parra and {Urry}, C. Megan},
        title = "{UGC 4211: A Confirmed Dual Active Galactic Nucleus in the Local Universe at 230 pc Nuclear Separation}",
      journal = {\apjl},
         year = 2023,
        month = jan,
       volume = {942},
       number = {1},
          eid = {L24},
        pages = {L24},
          doi = {10.3847/2041-8213/aca8f0},
archivePrefix = {arXiv},
       eprint = {2301.03609},
 primaryClass = {astro-ph.GA},
       adsurl = {https://ui.adsabs.harvard.edu/abs/2023ApJ...942L..24K}
}

@ARTICLE{Ricci_2017,
       author = {{Ricci}, C. and {Bauer}, F.~E. and {Treister}, E. and {Schawinski}, K. and {Privon}, G.~C. and {Blecha}, L. and {Arevalo}, P. and {Armus}, L. and {Harrison}, F. and {Ho}, L.~C. and {Iwasawa}, K. and {Sanders}, D.~B. and {Stern}, D.},
        title = "{Growing supermassive black holes in the late stages of galaxy mergers are heavily obscured}",
         year = 2017,
        month = jun,
       volume = {468},
       number = {2},
        pages = {1273-1299},
          doi = {10.1093/mnras/stx173},
archivePrefix = {arXiv},
       eprint = {1701.04825},
 primaryClass = {astro-ph.HE},
       adsurl = {https://ui.adsabs.harvard.edu/abs/2017MNRAS.468.1273R}
}

@ARTICLE{Ricci_2021,
       author = {{Ricci}, C. and {Privon}, G.~C. and {Pfeifle}, R.~W. and {Armus}, L. and {Iwasawa}, K. and {Torres-Alb{\`a}}, N. and {Satyapal}, S. and {Bauer}, F.~E. and {Treister}, E. and {Ho}, L.~C. and {Aalto}, S. and {Ar{\'e}valo}, P. and {Barcos-Mu{\~n}oz}, L. and {Charmandaris}, V. and {Diaz-Santos}, T. and {Evans}, A.~S. and {Gao}, T. and {Inami}, H. and {Koss}, M.~J. and {Lansbury}, G. and {Linden}, S.~T. and {Medling}, A. and {Sanders}, D.~B. and {Song}, Y. and {Stern}, D. and {U}, V. and {Ueda}, Y. and {Yamada}, S.},
        title = "{A hard X-ray view of luminous and ultra-luminous infrared galaxies in GOALS - I. AGN obscuration along the merger sequence}",
         year = 2021,
        month = oct,
       volume = {506},
       number = {4},
        pages = {5935-5950},
          doi = {10.1093/mnras/stab2052},
archivePrefix = {arXiv},
       eprint = {2107.10864},
 primaryClass = {astro-ph.GA},
       adsurl = {https://ui.adsabs.harvard.edu/abs/2021MNRAS.506.5935R}
}

@article{Blecha_2018,
	doi = {10.1093/mnras/sty1274},
  
	url = {https://doi.org/10.1093\%2Fmnras\%2Fsty1274},
  
	year = 2018,
	month = {may},
  
	publisher = {Oxford University Press ({OUP})},
  
	volume = {478},
  
	number = {3},
  
	pages = {3056--3071},
  
	author = {Laura Blecha and Gregory F Snyder and Shobita Satyapal and Sara L Ellison},
  
	title = {The power of infrared {AGN} selection in mergers: a theoretical study},
  
	journal = {Monthly Notices of the Royal Astronomical Society}
}

@ARTICLE{Hildebrand_1983,
       author = {{Hildebrand}, R.~H.},
        title = "{The determination of cloud masses and dust characteristics from submillimetre thermal emission.}",
         year = 1983,
        month = sep,
       volume = {24},
        pages = {267-282},
       adsurl = {https://ui.adsabs.harvard.edu/abs/1983QJRAS..24..267H}
}

@article{Oh_2018,
	doi = {10.3847/1538-4365/aaa7fd},
  
	url = {https://doi.org/10.3847\%2F1538-4365\%2Faaa7fd},
  
	year = 2018,
	month = {may},
  
	publisher = {American Astronomical Society},
  
	volume = {235},
  
	number = {1},
  
	pages = {4},
  
	author = {Kyuseok Oh and Michael Koss and Craig B. Markwardt and Kevin Schawinski and Wayne H. Baumgartner and Scott D. Barthelmy and S. Bradley Cenko and Neil Gehrels and Richard Mushotzky and Abigail Petulante and Claudio Ricci and Amy Lien and Benny Trakhtenbrot},
  
	title = {The 105-Month Swift-BAT All-sky Hard X-Ray Survey},
  
	journal = {The Astrophysical Journal Supplement Series}
}

@ARTICLE{Murphy_2012,
       author = {{Murphy}, E.~J. and {Bremseth}, J. and {Mason}, B.~S. and {Condon}, J.~J. and {Schinnerer}, E. and {Aniano}, G. and {Armus}, L. and {Helou}, G. and {Turner}, J.~L. and {Jarrett}, T.~H.},
        title = "{The Star Formation in Radio Survey: GBT 33 GHz Observations of Nearby Galaxy Nuclei and Extranuclear Star-forming Regions}",
      journal = {\apj},
         year = 2012,
        month = dec,
       volume = {761},
       number = {2},
          eid = {97},
        pages = {97},
          doi = {10.1088/0004-637X/761/2/97},
archivePrefix = {arXiv},
       eprint = {1210.3360},
 primaryClass = {astro-ph.CO},
       adsurl = {https://ui.adsabs.harvard.edu/abs/2012ApJ...761...97M}
}

@ARTICLE{Sanders_2003,
       author = {{Sanders}, D.~B. and {Mazzarella}, J.~M. and {Kim}, D. -C. and {Surace}, J.~A. and {Soifer}, B.~T.},
        title = "{The IRAS Revised Bright Galaxy Sample}",
      journal = {\aj},
         year = 2003,
        month = oct,
       volume = {126},
       number = {4},
        pages = {1607-1664},
          doi = {10.1086/376841},
archivePrefix = {arXiv},
       eprint = {astro-ph/0306263},
 primaryClass = {astro-ph},
       adsurl = {https://ui.adsabs.harvard.edu/abs/2003AJ....126.1607S}
}

@article{Bianchi_2008,
    author = {Bianchi, Stefano and Chiaberge, Marco and Piconcelli, Enrico and Guainazzi, Matteo and Matt, Giorgio},
    title = "{Chandra unveils a binary active galactic nucleus in Mrk 463}",
    journal = {Monthly Notices of the Royal Astronomical Society},
    volume = {386},
    number = {1},
    pages = {105-110},
    year = {2008},
    month = {03},
    issn = {0035-8711},
    doi = {10.1111/j.1365-2966.2008.13078.x},
    url = {https://doi.org/10.1111/j.1365-2966.2008.13078.x},
    eprint = {https://academic.oup.com/mnras/article-pdf/386/1/105/2995475/mnras0386-0105.pdf},
}

@ARTICLE{Paggi_2017,
       author = {{Paggi}, Alessandro and {Fabbiano}, Giuseppina and {Risaliti}, Guido and {Wang}, Junfeng and {Karovska}, Margarita and {Elvis}, Martin and {Maksym}, W. Peter and {McDowell}, Jonathan and {Gallagher}, Jay},
        title = "{X-Ray Emission from the Nuclear Region of Arp 220}",
      journal = {\apj},
         year = 2017,
        month = may,
       volume = {841},
       number = {1},
          eid = {44},
        pages = {44},
          doi = {10.3847/1538-4357/aa713b},
archivePrefix = {arXiv},
       eprint = {1705.01547},
 primaryClass = {astro-ph.GA},
       adsurl = {https://ui.adsabs.harvard.edu/abs/2017ApJ...841...44P}
}

@ARTICLE{Torres_Alba_2018,
       author = {{Torres-Alb{\`a}}, N. and {Iwasawa}, K. and {D{\'\i}az-Santos}, T. and {Charmandaris}, V. and {Ricci}, C. and {Chu}, J.~K. and {Sanders}, D.~B. and {Armus}, L. and {Barcos-Mu{\~n}oz}, L. and {Evans}, A.~S. and {Howell}, J.~H. and {Inami}, H. and {Linden}, S.~T. and {Medling}, A.~M. and {Privon}, G.~C. and {U}, V. and {Yoon}, I.},
        title = "{C-GOALS. II. Chandra observations of the lower luminosity sample of nearby luminous infrared galaxies in GOALS}",
      journal = {\aap},
         year = 2018,
        month = dec,
       volume = {620},
          eid = {A140},
        pages = {A140},
          doi = {10.1051/0004-6361/201834105},
archivePrefix = {arXiv},
       eprint = {1810.02371},
 primaryClass = {astro-ph.GA},
       adsurl = {https://ui.adsabs.harvard.edu/abs/2018A&A...620A.140T}
}

@ARTICLE{Kennicutt_1998,
       author = {{Kennicutt}, Robert C., Jr.},
        title = "{Star Formation in Galaxies Along the Hubble Sequence}",
      journal = {\araa},
         year = 1998,
        month = jan,
       volume = {36},
        pages = {189-232},
          doi = {10.1146/annurev.astro.36.1.189},
archivePrefix = {arXiv},
       eprint = {astro-ph/9807187},
 primaryClass = {astro-ph},
       adsurl = {https://ui.adsabs.harvard.edu/abs/1998ARA&A..36..189K}
}

@ARTICLE{Inoue_Doi_2018,
       author = {{Inoue}, Yoshiyuki and {Doi}, Akihiro},
        title = "{Detection of Coronal Magnetic Activity in nearby Active Supermassive Black Holes}",
      journal = {\apj},
         year = 2018,
        month = dec,
       volume = {869},
       number = {2},
          eid = {114},
        pages = {114},
          doi = {10.3847/1538-4357/aaeb95},
archivePrefix = {arXiv},
       eprint = {1810.10732},
 primaryClass = {astro-ph.HE},
       adsurl = {https://ui.adsabs.harvard.edu/abs/2018ApJ...869..114I}
}

@ARTICLE{Armus_2009,
       author = {{Armus}, L. and {Mazzarella}, J.~M. and {Evans}, A.~S. and {Surace}, J.~A. and {Sanders}, D.~B. and {Iwasawa}, K. and {Frayer}, D.~T. and {Howell}, J.~H. and {Chan}, B. and {Petric}, A. and {Vavilkin}, T. and {Kim}, D.~C. and {Haan}, S. and {Inami}, H. and {Murphy}, E.~J. and {Appleton}, P.~N. and {Barnes}, J.~E. and {Bothun}, G. and {Bridge}, C.~R. and {Charmandaris}, V. and {Jensen}, J.~B. and {Kewley}, L.~J. and {Lord}, S. and {Madore}, B.~F. and {Marshall}, J.~A. and {Melbourne}, J.~E. and {Rich}, J. and {Satyapal}, S. and {Schulz}, B. and {Spoon}, H.~W.~W. and {Sturm}, E. and {U}, V. and {Veilleux}, S. and {Xu}, K.},
        title = "{GOALS: The Great Observatories All-Sky LIRG Survey}",
      journal = {\pasp},
         year = 2009,
        month = jun,
       volume = {121},
       number = {880},
        pages = {559},
          doi = {10.1086/600092},
archivePrefix = {arXiv},
       eprint = {0904.4498},
 primaryClass = {astro-ph.CO},
       adsurl = {https://ui.adsabs.harvard.edu/abs/2009PASP..121..559A}
}

@article{Treister_2018,
   title={Optical, Near-IR, and Sub-mm IFU Observations of the Nearby Dual Active Galactic Nuclei MRK 463},
   volume={854},
   ISSN={1538-4357},
   url={http://dx.doi.org/10.3847/1538-4357/aaa963},
   DOI={10.3847/1538-4357/aaa963},
   number={2},
   journal={The Astrophysical Journal},
   publisher={American Astronomical Society},
   author={Treister, Ezequiel and Privon, George C. and Sartori, Lia F. and Nagar, Neil and Bauer, Franz E. and Schawinski, Kevin and Messias, Hugo and Ricci, Claudio and U, Vivian and Casey, Caitlin and Comerford, Julia M. and Muller-Sanchez, Francisco and Evans, Aaron S. and Finlez, Carolina and Koss, Michael and Sanders, David B. and Urry, C. Megan},
   year={2018},
   month=feb, pages={83} }

@ARTICLE{Treister_2020,
       author = {{Treister}, Ezequiel and {Messias}, Hugo and {Privon}, George C. and {Nagar}, Neil and {Medling}, Anne M. and {U}, Vivian and {Bauer}, Franz E. and {Cicone}, Claudia and {Mu{\~n}oz}, Loreto Barcos and {Evans}, Aaron S. and {Muller-Sanchez}, Francisco and {Comerford}, Julia M. and {Armus}, Lee and {Chang}, Chin-Shin and {Koss}, Michael and {Venturi}, Giacomo and {Schawinski}, Kevin and {Casey}, Caitlin and {Urry}, C. Megan and {Sanders}, David B. and {Scoville}, Nicholas and {Sheth}, Kartik},
        title = "{The Molecular Gas in the NGC 6240 Merging Galaxy System at the Highest Spatial Resolution}",
      journal = {\apj},
         year = 2020,
        month = feb,
       volume = {890},
       number = {2},
          eid = {149},
        pages = {149},
          doi = {10.3847/1538-4357/ab6b28},
archivePrefix = {arXiv},
       eprint = {2001.00601},
 primaryClass = {astro-ph.GA},
       adsurl = {https://ui.adsabs.harvard.edu/abs/2020ApJ...890..149T}
}

@ARTICLE{Mineo_2014,
       author = {{Mineo}, S. and {Gilfanov}, M. and {Lehmer}, B.~D. and {Morrison}, G.~E. and {Sunyaev}, R.},
        title = "{X-ray emission from star-forming galaxies - III. Calibration of the L$_{X}$-SFR relation up to redshift z {\ensuremath{\approx}} 1.3}",
      journal = {\mnras},
         year = 2014,
        month = jan,
       volume = {437},
       number = {2},
        pages = {1698-1707},
          doi = {10.1093/mnras/stt1999},
archivePrefix = {arXiv},
       eprint = {1207.2157},
 primaryClass = {astro-ph.HE},
       adsurl = {https://ui.adsabs.harvard.edu/abs/2014MNRAS.437.1698M}
}

@ARTICLE{Zhao_2021,
       author = {{Zhao}, X. and {Marchesi}, S. and {Ajello}, M. and {Cole}, D. and {Hu}, Z. and {Silver}, R. and {Torres-Alb{\`a}}, N.},
        title = "{The properties of the AGN torus as revealed from a set of unbiased NuSTAR observations}",
      journal = {\aap},
         year = 2021,
        month = jun,
       volume = {650},
          eid = {A57},
        pages = {A57},
          doi = {10.1051/0004-6361/202140297},
archivePrefix = {arXiv},
       eprint = {2011.03851},
 primaryClass = {astro-ph.GA},
       adsurl = {https://ui.adsabs.harvard.edu/abs/2021A&A...650A..57Z}
}

@ARTICLE{Behar_2018,
       author = {{Behar}, Ehud and {Vogel}, Stuart and {Baldi}, Ranieri D. and {Smith}, Krista L. and {Mushotzky}, Richard F.},
        title = "{The mm-wave compact component of an AGN}",
      journal = {\mnras},
         year = 2018,
        month = jul,
       volume = {478},
       number = {1},
        pages = {399-406},
          doi = {10.1093/mnras/sty850},
archivePrefix = {arXiv},
       eprint = {1803.06877},
 primaryClass = {astro-ph.GA},
       adsurl = {https://ui.adsabs.harvard.edu/abs/2018MNRAS.478..399B}
}

@ARTICLE{Sanders_1988,
       author = {{Sanders}, D.~B. and {Soifer}, B.~T. and {Elias}, J.~H. and {Madore}, B.~F. and {Matthews}, K. and {Neugebauer}, G. and {Scoville}, N.~Z.},
        title = "{Ultraluminous Infrared Galaxies and the Origin of Quasars}",
      journal = {\apj},
         year = 1988,
        month = feb,
       volume = {325},
        pages = {74},
          doi = {10.1086/165983},
       adsurl = {https://ui.adsabs.harvard.edu/abs/1988ApJ...325...74S}
}

@ARTICLE{Hopkins_2008,
       author = {{Hopkins}, Philip F. and {Hernquist}, Lars and {Cox}, Thomas J. and {Kere{\v{s}}}, Du{\v{s}}an},
        title = "{A Cosmological Framework for the Co-Evolution of Quasars, Supermassive Black Holes, and Elliptical Galaxies. I. Galaxy Mergers and Quasar Activity}",
      journal = {\apjs},
         year = 2008,
        month = apr,
       volume = {175},
       number = {2},
        pages = {356-389},
          doi = {10.1086/524362},
archivePrefix = {arXiv},
       eprint = {0706.1243},
 primaryClass = {astro-ph},
       adsurl = {https://ui.adsabs.harvard.edu/abs/2008ApJS..175..356H}
}

@INCOLLECTION{Lonsdale_2006,
       author = {{Lonsdale}, C.~J. and {Farrah}, D. and {Smith}, H.~E.},
        title = "{Ultraluminous Infrared Galaxies}",
    booktitle = {Astrophysics Update 2},
         year = 2006,
       editor = {{Mason}, John W.},
        pages = {285},
          doi = {10.1007/3-540-30313-8_9},
       adsurl = {https://ui.adsabs.harvard.edu/abs/2006asup.book..285L}
}

@ARTICLE{Sanders_1996,
  title     = "Luminous infrared galaxies",
  author    = "Sanders, D B and Mirabel, I F",
  journal   = "Annu. Rev. Astron. Astrophys.",
  publisher = "Annual Reviews",
  volume    =  34,
  number    =  1,
  pages     = "749--792",
  month     =  sep,
  year      =  1996,
  language  = "en"
}

@ARTICLE{Persic_Rephaeli_2007,
       author = {{Persic}, M. and {Rephaeli}, Y.},
        title = "{Galactic star formation rates gauged by stellar end-products}",
      journal = {\aap},
         year = 2007,
        month = feb,
       volume = {463},
       number = {2},
        pages = {481-492},
          doi = {10.1051/0004-6361:20054146},
archivePrefix = {arXiv},
       eprint = {astro-ph/0610321},
 primaryClass = {astro-ph},
       adsurl = {https://ui.adsabs.harvard.edu/abs/2007A&A...463..481P}
}

@ARTICLE{Lehmer_2010,
       author = {{Lehmer}, B.~D. and {Alexander}, D.~M. and {Bauer}, F.~E. and {Brandt}, W.~N. and {Goulding}, A.~D. and {Jenkins}, L.~P. and {Ptak}, A. and {Roberts}, T.~P.},
        title = "{A Chandra Perspective on Galaxy-wide X-ray Binary Emission and its Correlation with Star Formation Rate and Stellar Mass: New Results from Luminous Infrared Galaxies}",
      journal = {\apj},
         year = 2010,
        month = nov,
       volume = {724},
       number = {1},
        pages = {559-571},
          doi = {10.1088/0004-637X/724/1/559},
archivePrefix = {arXiv},
       eprint = {1009.3943},
 primaryClass = {astro-ph.CO},
       adsurl = {https://ui.adsabs.harvard.edu/abs/2010ApJ...724..559L}
}

@ARTICLE{Soifer_1984,
       author = {{Soifer}, B.~T. and {Helou}, G. and {Lonsdale}, C.~J. and {Neugebauer}, G. and {Hacking}, P. and {Houck}, J.~R. and {Low}, F.~J. and {Rice}, W. and {Rowan-Robinson}, M.},
        title = "{The remarkable infrared galaxy ARP 220 = IC 4553.}",
      journal = {\apjl},
         year = 1984,
        month = aug,
       volume = {283},
        pages = {L1-L4},
          doi = {10.1086/184319},
       adsurl = {https://ui.adsabs.harvard.edu/abs/1984ApJ...283L...1S}
}

@ARTICLE{Yun_2002,
       author = {{Yun}, Min S. and {Carilli}, C.~L.},
        title = "{Radio-to-Far-Infrared Spectral Energy Distribution and Photometric Redshifts for Dusty Starburst Galaxies}",
      journal = {\apj},
         year = 2002,
        month = mar,
       volume = {568},
       number = {1},
        pages = {88-98},
          doi = {10.1086/338924},
archivePrefix = {arXiv},
       eprint = {astro-ph/0112074},
 primaryClass = {astro-ph},
       adsurl = {https://ui.adsabs.harvard.edu/abs/2002ApJ...568...88Y}
}

@ARTICLE{Baumgartner_2013,
       author = {{Baumgartner}, W.~H. and {Tueller}, J. and {Markwardt}, C.~B. and {Skinner}, G.~K. and {Barthelmy}, S. and {Mushotzky}, R.~F. and {Evans}, P.~A. and {Gehrels}, N.},
        title = "{The 70 Month Swift-BAT All-sky Hard X-Ray Survey}",
      journal = {\apjs},
         year = 2013,
        month = aug,
       volume = {207},
       number = {2},
          eid = {19},
        pages = {19},
          doi = {10.1088/0067-0049/207/2/19},
archivePrefix = {arXiv},
       eprint = {1212.3336},
 primaryClass = {astro-ph.HE},
       adsurl = {https://ui.adsabs.harvard.edu/abs/2013ApJS..207...19B}
}

@ARTICLE{Jiang_2010,
       author = {{Jiang}, Yan-Fei and {Ciotti}, Luca and {Ostriker}, Jeremiah P. and {Spitkovsky}, Anatoly},
        title = "{Synchrotron Emission from Elliptical Galaxies Consequent to Active Galactic Nucleus Outbursts}",
      journal = {\apj},
         year = 2010,
        month = mar,
       volume = {711},
       number = {1},
        pages = {125-137},
          doi = {10.1088/0004-637X/711/1/125},
archivePrefix = {arXiv},
       eprint = {0904.4918},
 primaryClass = {astro-ph.CO},
       adsurl = {https://ui.adsabs.harvard.edu/abs/2010ApJ...711..125J}
}

@ARTICLE{Brinkmann_2000,
       author = {{Brinkmann}, W. and {Laurent-Muehleisen}, S.~A. and {Voges}, W. and {Siebert}, J. and {Becker}, R.~H. and {Brotherton}, M.~S. and {White}, R.~L. and {Gregg}, M.~D.},
        title = "{Radio and X-ray bright AGN: the ROSAT - FIRST correlation}",
      journal = {\aap},
         year = 2000,
        month = apr,
       volume = {356},
        pages = {445-462},
       adsurl = {https://ui.adsabs.harvard.edu/abs/2000A&A...356..445B}
}

@article{Salvato_2004,
doi = {10.1086/381692},
url = {https://dx.doi.org/10.1086/381692},
year = {2003},
month = {dec},
publisher = {},
volume = {600},
number = {1},
pages = {L31},
author = {Salvato, M. and Greiner, J. and Kuhlbrodt, B.},
title = {Multiwavelength Scaling Relations for Nuclei of Seyfert Galaxies*},
journal = {The Astrophysical Journal}
}

@article{Wang_2006,
doi = {10.1086/504401},
url = {https://dx.doi.org/10.1086/504401},
year = {2006},
month = {jul},
publisher = {},
volume = {645},
number = {2},
pages = {890},
author = {Wang, Ran and Wu, Xue-Bing and Kong, Min-Zhi},
title = {The Black Hole Fundamental Plane from a Uniform Sample of Radio and X-Ray-emitting Broad-Line AGNs},
journal = {The Astrophysical Journal}
}

@ARTICLE{Panessa_2007,
       author = {{Panessa}, F. and {Barcons}, X. and {Bassani}, L. and {Cappi}, M. and {Carrera}, F.~J. and {Ho}, L.~C. and {Pellegrini}, S.},
        title = "{The X-ray and radio connection in low-luminosity active nuclei}",
      journal = {\aap},
         year = 2007,
        month = may,
       volume = {467},
       number = {2},
        pages = {519-527},
          doi = {10.1051/0004-6361:20066943},
archivePrefix = {arXiv},
       eprint = {astro-ph/0701546},
 primaryClass = {astro-ph},
       adsurl = {https://ui.adsabs.harvard.edu/abs/2007A&A...467..519P}
}

@ARTICLE{Yamada_2021,
       author = {{Yamada}, Satoshi and {Ueda}, Yoshihiro and {Tanimoto}, Atsushi and {Imanishi}, Masatoshi and {Toba}, Yoshiki and {Ricci}, Claudio and {Privon}, George C.},
        title = "{Comprehensive Broadband X-Ray and Multiwavelength Study of Active Galactic Nuclei in 57 Local Luminous and Ultraluminous Infrared Galaxies Observed with NuSTAR and/or Swift/BAT}",
      journal = {\apjs},
         year = 2021,
        month = dec,
       volume = {257},
       number = {2},
          eid = {61},
        pages = {61},
          doi = {10.3847/1538-4365/ac17f5},
archivePrefix = {arXiv},
       eprint = {2107.10855},
 primaryClass = {astro-ph.GA},
       adsurl = {https://ui.adsabs.harvard.edu/abs/2021ApJS..257...61Y}
}

@INPROCEEDINGS{Becklin_1987,
       author = {{Becklin}, E.~E. and {Wynn-Williams}, C.~G.},
        title = "{Groundbased 1-MICRON to 32-MICRON Observations of ARP220 - Evidence for a Dust Embedded AGN / Active Galactic Nuclei}",
    booktitle = {NASA Conference Publication},
         year = 1987,
       editor = {{Lonsdale Persson}, Carol J.},
       series = {NASA Conference Publication},
       volume = {2466},
        month = may,
        pages = {643-650},
       adsurl = {https://ui.adsabs.harvard.edu/abs/1987NASCP2466..643B}
}

@ARTICLE{Iwasawa_2005,
       author = {{Iwasawa}, K. and {Sanders}, D.~B. and {Evans}, A.~S. and {Trentham}, N. and {Miniutti}, G. and {Spoon}, H.~W.~W.},
        title = "{Fe K emission in the ultraluminous infrared galaxy Arp 220}",
      journal = {\mnras},
         year = 2005,
        month = feb,
       volume = {357},
       number = {2},
        pages = {565-571},
          doi = {10.1111/j.1365-2966.2005.08644.x},
archivePrefix = {arXiv},
       eprint = {astro-ph/0411562},
 primaryClass = {astro-ph},
       adsurl = {https://ui.adsabs.harvard.edu/abs/2005MNRAS.357..565I}
}

@ARTICLE{Barcos-Munoz_2015,
       author = {{Barcos-Mu{\~n}oz}, L. and {Leroy}, A.~K. and {Evans}, A.~S. and {Privon}, G.~C. and {Armus}, L. and {Condon}, J. and {Mazzarella}, J.~M. and {Meier}, D.~S. and {Momjian}, E. and {Murphy}, E.~J. and {Ott}, J. and {Reichardt}, A. and {Sakamoto}, K. and {Sanders}, D.~B. and {Schinnerer}, E. and {Stierwalt}, S. and {Surace}, J.~A. and {Thompson}, T.~A. and {Walter}, F.},
        title = "{High-resolution Radio Continuum Measurements of the Nuclear Disks of Arp 220}",
      journal = {\apj},
         year = 2015,
        month = jan,
       volume = {799},
       number = {1},
          eid = {10},
        pages = {10},
          doi = {10.1088/0004-637X/799/1/10},
archivePrefix = {arXiv},
       eprint = {1411.0932},
 primaryClass = {astro-ph.GA},
       adsurl = {https://ui.adsabs.harvard.edu/abs/2015ApJ...799...10B}
}

@ARTICLE{Perna_2024,
       author = {{Perna}, Michele and {Arribas}, Santiago and {Lamperti}, Isabella and {Pereira-Santaella}, Miguel and {Ulivi}, Lorenzo and {B{\"o}ker}, Torsten and {Maiolino}, Roberto and {Bunker}, Andrew J. and {Charlot}, St{\'e}phane and {Cresci}, Giovanni and {Rodr{\'\i}guez Del Pino}, Bruno and {D'Eugenio}, Francesco and {{\"U}bler}, Hannah and {Fahrion}, Katja and {Ceci}, Matteo},
        title = "{No evidence of active galactic nucleus features in the nuclei of Arp 220 from JWST/NIRSpec IFS}",
      journal = {\aap},
         year = 2024,
        month = oct,
       volume = {690},
          eid = {A171},
        pages = {A171},
          doi = {10.1051/0004-6361/202450094},
archivePrefix = {arXiv},
       eprint = {2403.13948},
 primaryClass = {astro-ph.GA},
       adsurl = {https://ui.adsabs.harvard.edu/abs/2024A&A...690A.171P}
}

@ARTICLE{CasaTeam_2022,
       author = {{CASA Team} and {Bean}, Ben and {Bhatnagar}, Sanjay and {Castro}, Sandra and {Donovan Meyer}, Jennifer and {Emonts}, Bjorn and {Garcia}, Enrique and {Garwood}, Robert and {Golap}, Kumar and {Gonzalez Villalba}, Justo and {Harris}, Pamela and {Hayashi}, Yohei and {Hoskins}, Josh and {Hsieh}, Mingyu and {Jagannathan}, Preshanth and {Kawasaki}, Wataru and {Keimpema}, Aard and {Kettenis}, Mark and {Lopez}, Jorge and {Marvil}, Joshua and {Masters}, Joseph and {McNichols}, Andrew and {Mehringer}, David and {Miel}, Renaud and {Moellenbrock}, George and {Montesino}, Federico and {Nakazato}, Takeshi and {Ott}, Juergen and {Petry}, Dirk and {Pokorny}, Martin and {Raba}, Ryan and {Rau}, Urvashi and {Schiebel}, Darrell and {Schweighart}, Neal and {Sekhar}, Srikrishna and {Shimada}, Kazuhiko and {Small}, Des and {Steeb}, Jan-Willem and {Sugimoto}, Kanako and {Suoranta}, Ville and {Tsutsumi}, Takahiro and {van Bemmel}, Ilse M. and {Verkouter}, Marjolein and {Wells}, Akeem and {Xiong}, Wei and {Szomoru}, Arpad and {Griffith}, Morgan and {Glendenning}, Brian and {Kern}, Jeff},
        title = "{CASA, the Common Astronomy Software Applications for Radio Astronomy}",
      journal = {\pasp},
         year = 2022,
        month = nov,
       volume = {134},
       number = {1041},
          eid = {114501},
        pages = {114501},
          doi = {10.1088/1538-3873/ac9642},
archivePrefix = {arXiv},
       eprint = {2210.02276},
 primaryClass = {astro-ph.IM},
       adsurl = {https://ui.adsabs.harvard.edu/abs/2022PASP..134k4501C}
}

@ARTICLE{Mineo_2012,
       author = {{Mineo}, S. and {Gilfanov}, M. and {Sunyaev}, R.},
        title = "{X-ray emission from star-forming galaxies - II. Hot interstellarmedium}",
      journal = {\mnras},
         year = 2012,
        month = nov,
       volume = {426},
       number = {3},
        pages = {1870-1883},
          doi = {10.1111/j.1365-2966.2012.21831.x},
archivePrefix = {arXiv},
       eprint = {1205.3715},
 primaryClass = {astro-ph.HE},
       adsurl = {https://ui.adsabs.harvard.edu/abs/2012MNRAS.426.1870M}
}

@ARTICLE{Ricci_2017b,
       author = {{Ricci}, C. and {Trakhtenbrot}, B. and {Koss}, M.~J. and {Ueda}, Y. and {Delvecchio}, I. and {Treister}, E. and {Schawinski}, K. and {Paltani}, S. and {Oh}, K. and {Lamperti}, I. and {Berney}, S. and {Gandhi}, P. and {Ichikawa}, K. and {Bauer}, F.~E. and {Ho}, L.~C. and {Asmus}, D. and {Beckmann}, V. and {Soldi}, S. and {Balokovi{\'c}}, M. and {Gehrels}, N. and {Markwardt}, C.~B.},
        title = "{BAT AGN Spectroscopic Survey. V. X-Ray Properties of the Swift/BAT 70-month AGN Catalog}",
      journal = {\apjs},
         year = 2017,
        month = dec,
       volume = {233},
       number = {2},
          eid = {17},
        pages = {17},
          doi = {10.3847/1538-4365/aa96ad},
archivePrefix = {arXiv},
       eprint = {1709.03989},
 primaryClass = {astro-ph.HE},
       adsurl = {https://ui.adsabs.harvard.edu/abs/2017ApJS..233...17R}
}

@ARTICLE{Done_2007,
       author = {{Done}, Chris and {Gierli{\'n}ski}, Marek and {Kubota}, Aya},
        title = "{Modelling the behaviour of accretion flows in X-ray binaries. Everything you always wanted to know about accretion but were afraid to ask}",
      journal = {\aapr},
         year = 2007,
        month = dec,
       volume = {15},
       number = {1},
        pages = {1-66},
          doi = {10.1007/s00159-007-0006-1},
archivePrefix = {arXiv},
       eprint = {0708.0148},
 primaryClass = {astro-ph},
       adsurl = {https://ui.adsabs.harvard.edu/abs/2007A&ARv..15....1D}
}

@ARTICLE{Iwasawa_2001,
       author = {{Iwasawa}, K. and {Matt}, G. and {Guainazzi}, M. and {Fabian}, A.~C.},
        title = "{A hard X-ray constraint on the presence of an AGN in the ultraluminous infrared galaxy Arp 220}",
      journal = {\mnras},
         year = 2001,
        month = sep,
       volume = {326},
       number = {3},
        pages = {894-900},
          doi = {10.1046/j.1365-8711.2001.04478.x},
archivePrefix = {arXiv},
       eprint = {astro-ph/0103417},
 primaryClass = {astro-ph},
       adsurl = {https://ui.adsabs.harvard.edu/abs/2001MNRAS.326..894I}
}

@ARTICLE{Teng_2015,
       author = {{Teng}, Stacy H. and {Rigby}, Jane R. and {Stern}, Daniel and {Ptak}, Andrew and {Alexander}, D.~M. and {Bauer}, Franz E. and {Boggs}, Stephen E. and {Brandt}, W. Niel and {Christensen}, Finn E. and {Comastri}, Andrea and {Craig}, William W. and {Farrah}, Duncan and {Gandhi}, Poshak and {Hailey}, Charles J. and {Harrison}, Fiona A. and {Hickox}, Ryan C. and {Koss}, Michael and {Luo}, Bin and {Treister}, Ezequiel and {Zhang}, William W.},
        title = "{A NuSTAR Survey of Nearby Ultraluminous Infrared Galaxies}",
      journal = {\apj},
         year = 2015,
        month = nov,
       volume = {814},
       number = {1},
          eid = {56},
        pages = {56},
          doi = {10.1088/0004-637X/814/1/56},
archivePrefix = {arXiv},
       eprint = {1510.04453},
 primaryClass = {astro-ph.GA},
       adsurl = {https://ui.adsabs.harvard.edu/abs/2015ApJ...814...56T}
}

@dataset{Konig_2016,
       author = {{Koenig}, S. and {Garcia-Marin}, M. and {Eckart}, A. and {Downes}, D. and {Scharwachter}, J.},
        title = "{VizieR Online Data Catalog: Arp 220 CO (2-1) and (1-0) images (Koenig+, 2012)}",
 howpublished = {VizieR On-line Data Catalog: J/ApJ/754/58. Originally published in: 2012ApJ...754...58K},
         year = 2016,
        month = jul,
          eid = {J/ApJ/754/58},
       adsurl = {https://ui.adsabs.harvard.edu/abs/2016yCat..17540058K}
}

@ARTICLE{Sazonov_2017,
       author = {{Sazonov}, S. and {Khabibullin}, I.},
        title = "{The intrinsic collective X-ray spectrum of luminous high-mass X-ray binaries}",
      journal = {\mnras},
         year = 2017,
        month = jun,
       volume = {468},
       number = {2},
        pages = {2249-2255},
          doi = {10.1093/mnras/stx626},
archivePrefix = {arXiv},
       eprint = {1703.03354},
 primaryClass = {astro-ph.HE},
       adsurl = {https://ui.adsabs.harvard.edu/abs/2017MNRAS.468.2249S}
}

@ARTICLE{Seifina_2016,
       author = {{Seifina}, Elena and {Titarchuk}, Lev and {Shaposhnikov}, Nikolai},
        title = "{X-Ray Spectra of the High-mass X-Ray Binary 4U 1700-37 Using BeppoSAX, Suzaku, and RXTE Observations}",
      journal = {\apj},
         year = 2016,
        month = apr,
       volume = {821},
       number = {1},
          eid = {23},
        pages = {23},
          doi = {10.3847/0004-637X/821/1/23},
archivePrefix = {arXiv},
       eprint = {1603.05504},
 primaryClass = {astro-ph.HE},
       adsurl = {https://ui.adsabs.harvard.edu/abs/2016ApJ...821...23S}
}

@ARTICLE{Condon_1992,
       author = {{Condon}, J.~J.},
        title = "{Radio emission from normal galaxies.}",
      journal = {\araa},
         year = 1992,
        month = jan,
       volume = {30},
        pages = {575-611},
          doi = {10.1146/annurev.aa.30.090192.003043},
       adsurl = {https://ui.adsabs.harvard.edu/abs/1992ARA&A..30..575C}
}

@ARTICLE{Iwasawa_2011,
       author = {{Iwasawa}, K. and {Sanders}, D.~B. and {Teng}, S.~H. and {U}, Vivian and {Armus}, L. and {Evans}, A.~S. and {Howell}, J.~H. and {Komossa}, S. and {Mazzarella}, J.~M. and {Petric}, A.~O. and {Surace}, J.~A. and {Vavilkin}, T. and {Veilleux}, S. and {Trentham}, N.},
        title = "{C-GOALS: Chandra observations of a complete sample of luminous infrared galaxies from the IRAS Revised Bright Galaxy Survey}",
      journal = {\aap},
         year = 2011,
        month = may,
       volume = {529},
          eid = {A106},
        pages = {A106},
          doi = {10.1051/0004-6361/201015264},
archivePrefix = {arXiv},
       eprint = {1103.2755},
 primaryClass = {astro-ph.CO},
       adsurl = {https://ui.adsabs.harvard.edu/abs/2011A&A...529A.106I}
}

@article{Piro_2023,
    author = {Piro, L and Colpi, M and Aird, J and Mangiagli, A and Fabian, A C and Guainazzi, M and Marsat, S and Sesana, A and McNamara, P and Bonetti, M and Rossi, E M and Tanvir, N R and Baker, J G and Belanger, G and Dal Canton, T and Jennrich, O and Katz, M L and Luetzgendorf, N},
    title = {Chasing supermassive black hole merging events with Athena and LISA},
    journal = {Monthly Notices of the Royal Astronomical Society},
    volume = {521},
    number = {2},
    pages = {2577-2592},
    year = {2023},
    month = {03},
    issn = {0035-8711},
    doi = {10.1093/mnras/stad659},
    url = {https://doi.org/10.1093/mnras/stad659},
    eprint = {https://academic.oup.com/mnras/article-pdf/521/2/2577/49570717/stad659.pdf},
}

@ARTICLE{Nandra_2013,
       author = {{Nandra}, Kirpal and {Barret}, Didier and {Barcons}, Xavier and {Fabian}, Andy and {den Herder}, Jan-Willem and {Piro}, Luigi and {Watson}, Mike and {Adami}, Christophe and {Aird}, James and {Afonso}, Jose Manuel and {Alexander}, Dave and {Argiroffi}, Costanza and {Amati}, Lorenzo and {Arnaud}, Monique and {Atteia}, Jean-Luc and {Audard}, Marc and {Badenes}, Carles and {Ballet}, Jean and {Ballo}, Lucia and {Bamba}, Aya and {Bhardwaj}, Anil and {Stefano Battistelli}, Elia and {Becker}, Werner and {De Becker}, Micha{\"e}l and {Behar}, Ehud and {Bianchi}, Stefano and {Biffi}, Veronica and {B{\^\i}rzan}, Laura and {Bocchino}, Fabrizio and {Bogdanov}, Slavko and {Boirin}, Laurence and {Boller}, Thomas and {Borgani}, Stefano and {Borm}, Katharina and {Bouch{\'e}}, Nicolas and {Bourdin}, Herv{\'e} and {Bower}, Richard and {Braito}, Valentina and {Branchini}, Enzo and {Branduardi-Raymont}, Graziella and {Bregman}, Joel and {Brenneman}, Laura and {Brightman}, Murray and {Br{\"u}ggen}, Marcus and {Buchner}, Johannes and {Bulbul}, Esra and {Brusa}, Marcella and {Bursa}, Michal and {Caccianiga}, Alessandro and {Cackett}, Ed and {Campana}, Sergio and {Cappelluti}, Nico and {Cappi}, Massimo and {Carrera}, Francisco and {Ceballos}, Maite and {Christensen}, Finn and {Chu}, You-Hua and {Churazov}, Eugene and {Clerc}, Nicolas and {Corbel}, Stephane and {Corral}, Amalia and {Comastri}, Andrea and {Costantini}, Elisa and {Croston}, Judith and {Dadina}, Mauro and {D'Ai}, Antonino and {Decourchelle}, Anne and {Della Ceca}, Roberto and {Dennerl}, Konrad and {Dolag}, Klaus and {Done}, Chris and {Dovciak}, Michal and {Drake}, Jeremy and {Eckert}, Dominique and {Edge}, Alastair and {Ettori}, Stefano and {Ezoe}, Yuichiro and {Feigelson}, Eric and {Fender}, Rob and {Feruglio}, Chiara and {Finoguenov}, Alexis and {Fiore}, Fabrizio and {Galeazzi}, Massimiliano and {Gallagher}, Sarah and {Gandhi}, Poshak and {Gaspari}, Massimo and {Gastaldello}, Fabio and {Georgakakis}, Antonis and {Georgantopoulos}, Ioannis and {Gilfanov}, Marat and {Gitti}, Myriam and {Gladstone}, Randy and {Goosmann}, Rene and {Gosset}, Eric and {Grosso}, Nicolas and {Guedel}, Manuel and {Guerrero}, Martin and {Haberl}, Frank and {Hardcastle}, Martin and {Heinz}, Sebastian and {Alonso Herrero}, Almudena and {Herv{\'e}}, Anthony and {Holmstrom}, Mats and {Iwasawa}, Kazushi and {Jonker}, Peter and {Kaastra}, Jelle and {Kara}, Erin and {Karas}, Vladimir and {Kastner}, Joel and {King}, Andrew and {Kosenko}, Daria and {Koutroumpa}, Dimita and {Kraft}, Ralph and {Kreykenbohm}, Ingo and {Lallement}, Rosine and {Lanzuisi}, Giorgio and {Lee}, J. and {Lemoine-Goumard}, Marianne and {Lobban}, Andrew and {Lodato}, Giuseppe and {Lovisari}, Lorenzo and {Lotti}, Simone and {McCharthy}, Ian and {McNamara}, Brian and {Maggio}, Antonio and {Maiolino}, Roberto and {De Marco}, Barbara and {de Martino}, Domitilla and {Mateos}, Silvia and {Matt}, Giorgio and {Maughan}, Ben and {Mazzotta}, Pasquale and {Mendez}, Mariano and {Merloni}, Andrea and {Micela}, Giuseppina and {Miceli}, Marco and {Mignani}, Robert and {Miller}, Jon and {Miniutti}, Giovanni and {Molendi}, Silvano and {Montez}, Rodolfo and {Moretti}, Alberto and {Motch}, Christian and {Naz{\'e}}, Ya{\"e}l and {Nevalainen}, Jukka and {Nicastro}, Fabrizio and {Nulsen}, Paul and {Ohashi}, Takaya and {O'Brien}, Paul and {Osborne}, Julian and {Oskinova}, Lida and {Pacaud}, Florian and {Paerels}, Frederik and {Page}, Mat and {Papadakis}, Iossif and {Pareschi}, Giovanni and {Petre}, Robert and {Petrucci}, Pierre-Olivier and {Piconcelli}, Enrico and {Pillitteri}, Ignazio and {Pinto}, C. and {de Plaa}, Jelle and {Pointecouteau}, Etienne and {Ponman}, Trevor and {Ponti}, Gabriele and {Porquet}, Delphine and {Pounds}, Ken and {Pratt}, Gabriel and {Predehl}, Peter and {Proga}, Daniel and {Psaltis}, Dimitrios and {Rafferty}, David and {Ramos-Ceja}, Miriam and {Ranalli}, Piero and {Rasia}, Elena and {Rau}, Arne and {Rauw}, Gregor and {Rea}, Nanda and {Read}, Andy and {Reeves}, James and {Reiprich}, Thomas and {Renaud}, Matthieu and {Reynolds}, Chris and {Risaliti}, Guido and {Rodriguez}, Jerome and {Rodriguez Hidalgo}, Paola and {Roncarelli}, Mauro and {Rosario}, David and {Rossetti}, Mariachiara and {Rozanska}, Agata and {Rovilos}, Emmanouil and {Salvaterra}, Ruben and {Salvato}, Mara and {Di Salvo}, Tiziana and {Sanders}, Jeremy and {Sanz-Forcada}, Jorge and {Schawinski}, Kevin and {Schaye}, Joop and {Schwope}, Axel and {Sciortino}, Salvatore},
        title = "{The Hot and Energetic Universe: A White Paper presenting the science theme motivating the Athena+ mission}",
      journal = {arXiv e-prints},
         year = 2013,
        month = jun,
          eid = {arXiv:1306.2307},
        pages = {arXiv:1306.2307},
          doi = {10.48550/arXiv.1306.2307},
archivePrefix = {arXiv},
       eprint = {1306.2307},
 primaryClass = {astro-ph.HE},
       adsurl = {https://ui.adsabs.harvard.edu/abs/2013arXiv1306.2307N}
}

@ARTICLE{Amaro-Seoane_2017,
       author = {{Amaro-Seoane}, Pau and {Audley}, Heather and {Babak}, Stanislav and {Baker}, John and {Barausse}, Enrico and {Bender}, Peter and {Berti}, Emanuele and {Binetruy}, Pierre and {Born}, Michael and {Bortoluzzi}, Daniele and {Camp}, Jordan and {Caprini}, Chiara and {Cardoso}, Vitor and {Colpi}, Monica and {Conklin}, John and {Cornish}, Neil and {Cutler}, Curt and {Danzmann}, Karsten and {Dolesi}, Rita and {Ferraioli}, Luigi and {Ferroni}, Valerio and {Fitzsimons}, Ewan and {Gair}, Jonathan and {Gesa Bote}, Lluis and {Giardini}, Domenico and {Gibert}, Ferran and {Grimani}, Catia and {Halloin}, Hubert and {Heinzel}, Gerhard and {Hertog}, Thomas and {Hewitson}, Martin and {Holley-Bockelmann}, Kelly and {Hollington}, Daniel and {Hueller}, Mauro and {Inchauspe}, Henri and {Jetzer}, Philippe and {Karnesis}, Nikos and {Killow}, Christian and {Klein}, Antoine and {Klipstein}, Bill and {Korsakova}, Natalia and {Larson}, Shane L and {Livas}, Jeffrey and {Lloro}, Ivan and {Man}, Nary and {Mance}, Davor and {Martino}, Joseph and {Mateos}, Ignacio and {McKenzie}, Kirk and {McWilliams}, Sean T and {Miller}, Cole and {Mueller}, Guido and {Nardini}, Germano and {Nelemans}, Gijs and {Nofrarias}, Miquel and {Petiteau}, Antoine and {Pivato}, Paolo and {Plagnol}, Eric and {Porter}, Ed and {Reiche}, Jens and {Robertson}, David and {Robertson}, Norna and {Rossi}, Elena and {Russano}, Giuliana and {Schutz}, Bernard and {Sesana}, Alberto and {Shoemaker}, David and {Slutsky}, Jacob and {Sopuerta}, Carlos F. and {Sumner}, Tim and {Tamanini}, Nicola and {Thorpe}, Ira and {Troebs}, Michael and {Vallisneri}, Michele and {Vecchio}, Alberto and {Vetrugno}, Daniele and {Vitale}, Stefano and {Volonteri}, Marta and {Wanner}, Gudrun and {Ward}, Harry and {Wass}, Peter and {Weber}, William and {Ziemer}, John and {Zweifel}, Peter},
        title = "{Laser Interferometer Space Antenna}",
      journal = {arXiv e-prints},
         year = 2017,
        month = feb,
          eid = {arXiv:1702.00786},
        pages = {arXiv:1702.00786},
          doi = {10.48550/arXiv.1702.00786},
archivePrefix = {arXiv},
       eprint = {1702.00786},
 primaryClass = {astro-ph.IM},
       adsurl = {https://ui.adsabs.harvard.edu/abs/2017arXiv170200786A}
}

@Inbook{Fornasini_2022,
author="Fornasini, Francesca
and Antoniou, Vallia
and Dubus, Guillaume",
editor="Bambi, Cosimo
and Santangelo, Andrea",
title="High-Mass X-ray Binaries",
bookTitle="Handbook of X-ray and Gamma-ray Astrophysics",
year="2022",
publisher="Springer Nature Singapore",
address="Singapore",
pages="1--55",
isbn="978-981-16-4544-0",
doi="10.1007/978-981-16-4544-0_95-1",
url="https://doi.org/10.1007/978-981-16-4544-0_95-1"
}

@ARTICLE{Lightman_1988,
       author = {{Lightman}, Alan P. and {White}, Timothy R.},
        title = "{Effects of Cold Matter in Active Galactic Nuclei: A Broad Hump in the X-Ray Spectra}",
      journal = {\apj},
         year = 1988,
        month = dec,
       volume = {335},
        pages = {57},
          doi = {10.1086/166905},
       adsurl = {https://ui.adsabs.harvard.edu/abs/1988ApJ...335...57L}
}

@ARTICLE{George_1991,
       author = {{George}, I.~M. and {Fabian}, A.~C.},
        title = "{X-ray reflection from cold matter in Active Galactic Nuclei and X-ray binaries.}",
      journal = {\mnras},
         year = 1991,
        month = mar,
       volume = {249},
        pages = {352},
          doi = {10.1093/mnras/249.2.352},
       adsurl = {https://ui.adsabs.harvard.edu/abs/1991MNRAS.249..352G}
}

@misc{zhang_1996,
      title={High Energy Continuum Spectra from X-Ray Binaries}, 
      author={S. N. Zhang},
      year={1996},
      eprint={astro-ph/9611039},
      archivePrefix={arXiv},
      primaryClass={astro-ph},
      url={https://arxiv.org/abs/astro-ph/9611039}, 
}

@ARTICLE{Coburn_2002,
       author = {{Coburn}, W. and {Heindl}, W.~A. and {Rothschild}, R.~E. and {Gruber}, D.~E. and {Kreykenbohm}, I. and {Wilms}, J. and {Kretschmar}, P. and {Staubert}, R.},
        title = "{Magnetic Fields of Accreting X-Ray Pulsars with the Rossi X-Ray Timing Explorer}",
      journal = {\apj},
         year = 2002,
        month = nov,
       volume = {580},
       number = {1},
        pages = {394-412},
          doi = {10.1086/343033},
archivePrefix = {arXiv},
       eprint = {astro-ph/0207325},
 primaryClass = {astro-ph},
       adsurl = {https://ui.adsabs.harvard.edu/abs/2002ApJ...580..394C}
}

@article{Koss_2016,
   title={A NEW POPULATION OF COMPTON-THICK AGNs IDENTIFIED USING THE SPECTRAL CURVATURE ABOVE 10 keV},
   volume={825},
   ISSN={1538-4357},
   url={http://dx.doi.org/10.3847/0004-637X/825/2/85},
   DOI={10.3847/0004-637x/825/2/85},
   number={2},
   journal={The Astrophysical Journal},
   publisher={American Astronomical Society},
   author={Koss, Michael J. and Assef, R. and Baloković, M. and Stern, D. and Gandhi, P. and Lamperti, I. and Alexander, D. M. and Ballantyne, D. R. and Bauer, F. E. and Berney, S. and Brandt, W. N. and Comastri, A. and Gehrels, N. and Harrison, F. A. and Lansbury, G. and Markwardt, C. and Ricci, C. and Rivers, E. and Schawinski, K. and Trakhtenbrot, B. and Treister, E. and Urry, C. Megan},
   year={2016},
   month=jul, pages={85} }

@ARTICLE{Koss_2017,
       author = {{Koss}, Michael and {Trakhtenbrot}, Benny and {Ricci}, Claudio and {Lamperti}, Isabella and {Oh}, Kyuseok and {Berney}, Simon and {Schawinski}, Kevin and {Balokovi{\'c}}, Mislav and {Baronchelli}, Linda and {Crenshaw}, D. Michael and {Fischer}, Travis and {Gehrels}, Neil and {Harrison}, Fiona and {Hashimoto}, Yasuhiro and {Hogg}, Drew and {Ichikawa}, Kohei and {Masetti}, Nicola and {Mushotzky}, Richard and {Sartori}, Lia and {Stern}, Daniel and {Treister}, Ezequiel and {Ueda}, Yoshihiro and {Veilleux}, Sylvain and {Winter}, Lisa},
        title = "{BAT AGN Spectroscopic Survey. I. Spectral Measurements, Derived Quantities, and AGN Demographics}",
      journal = {\apj},
         year = 2017,
        month = nov,
       volume = {850},
       number = {1},
          eid = {74},
        pages = {74},
          doi = {10.3847/1538-4357/aa8ec9},
archivePrefix = {arXiv},
       eprint = {1707.08123},
 primaryClass = {astro-ph.HE},
       adsurl = {https://ui.adsabs.harvard.edu/abs/2017ApJ...850...74K}
}

@article{Koss_2018,
   title={A population of luminous accreting black holes with hidden mergers},
   volume={563},
   ISSN={1476-4687},
   url={http://dx.doi.org/10.1038/s41586-018-0652-7},
   DOI={10.1038/s41586-018-0652-7},
   number={7730},
   journal={Nature},
   publisher={Springer Science and Business Media LLC},
   author={Koss, Michael J. and Blecha, Laura and Bernhard, Phillip and Hung, Chao-Ling and Lu, Jessica R. and Trakhtenbrot, Benny and Treister, Ezequiel and Weigel, Anna and Sartori, Lia F. and Mushotzky, Richard and Schawinski, Kevin and Ricci, Claudio and Veilleux, Sylvain and Sanders, David B.},
   year={2018},
   month=nov, pages={214–216} }

@article{Koss_2012,
   title={UNDERSTANDING DUAL ACTIVE GALACTIC NUCLEUS ACTIVATION IN THE NEARBY UNIVERSE},
   volume={746},
   ISSN={2041-8213},
   url={http://dx.doi.org/10.1088/2041-8205/746/2/L22},
   DOI={10.1088/2041-8205/746/2/l22},
   number={2},
   journal={The Astrophysical Journal},
   publisher={American Astronomical Society},
   author={Koss, Michael and Mushotzky, Richard and Treister, Ezequiel and Veilleux, Sylvain and Vasudevan, Ranjan and Trippe, Margaret},
   year={2012},
   month=feb, pages={L22} }

@ARTICLE{Boorman_2025,
       author = {{Boorman}, Peter G. and {Gandhi}, Poshak and {Buchner}, Johannes and {Stern}, Daniel and {Ricci}, Claudio and {Balokovi{\'c}}, Mislav and {Asmus}, Daniel and {Harrison}, Fiona A. and {Svoboda}, Ji{\v{r}}{\'\i} and {Greenwell}, Claire and {Koss}, Michael J. and {Alexander}, David M. and {Annuar}, Adlyka and {Bauer}, Franz E. and {Brandt}, William N. and {Brightman}, Murray and {Civano}, Francesca and {Chen}, Chien-Ting J. and {Farrah}, Duncan and {Forster}, Karl and {Grefenstette}, Brian and {H{\"o}nig}, Sebastian F. and {Hill}, Adam B. and {Kammoun}, Elias and {Lansbury}, George and {Lanz}, Lauranne and {LaMassa}, Stephanie and {Madsen}, Kristin and {Marchesi}, Stefano and {Middleton}, Matthew and {Mingo}, Beatriz and {Parker}, Michael L. and {Treister}, Ezequiel and {Ueda}, Yoshihiro and {Urry}, C. Megan and {Zappacosta}, Luca},
        title = "{The NuSTAR Local AGN N $_{H}$ Distribution Survey (NuLANDS). I. Toward a Truly Representative Column Density Distribution in the Local Universe}",
      journal = {\apj},
         year = 2025,
        month = jan,
       volume = {978},
       number = {1},
          eid = {118},
        pages = {118},
          doi = {10.3847/1538-4357/ad8236},
archivePrefix = {arXiv},
       eprint = {2410.07339},
 primaryClass = {astro-ph.GA},
       adsurl = {https://ui.adsabs.harvard.edu/abs/2025ApJ...978..118B}
}

@article{Koss_2011,
   title={CHANDRA
                    DISCOVERY OF A BINARY ACTIVE GALACTIC NUCLEUS IN Mrk 739},
   volume={735},
   ISSN={2041-8213},
   url={http://dx.doi.org/10.1088/2041-8205/735/2/L42},
   DOI={10.1088/2041-8205/735/2/l42},
   number={2},
   journal={The Astrophysical Journal},
   publisher={American Astronomical Society},
   author={Koss, Michael and Mushotzky, Richard and Veilleux, Sylvain and Vasudevan, Ranjan and Miller, Neal and Sanders, D. B. and Schawinski, Kevin and Trippe, Margaret},
   year={2011},
   month=jun, pages={L42} }

@article{Koss_2010,
   title={MERGING AND CLUSTERING OF THE
                    SWIFT
                    BAT AGN SAMPLE},
   volume={716},
   ISSN={2041-8213},
   url={http://dx.doi.org/10.1088/2041-8205/716/2/L125},
   DOI={10.1088/2041-8205/716/2/l125},
   number={2},
   journal={The Astrophysical Journal},
   publisher={American Astronomical Society},
   author={Koss, Michael and Mushotzky, Richard and Veilleux, Sylvain and Winter, Lisa},
   year={2010},
   month=may, pages={L125–L130} }

@misc{Pfeifle_2024,
      title={Super-Size Me: The Big Multi-AGN Catalog (The Big MAC), Data Release 1: The Source Catalog}, 
      author={Ryan W. Pfeifle and Kimberly A. Weaver and Nathan J. Secrest and Barry Rothberg and David R. Patton},
      year={2024},
      eprint={2411.12799},
      archivePrefix={arXiv},
      primaryClass={astro-ph.GA},
      url={https://arxiv.org/abs/2411.12799}, 
}

@ARTICLE{Franceschini_2003,
       author = {{Franceschini}, A. and {Braito}, V. and {Persic}, M. and {Della Ceca}, R. and {Bassani}, L. and {Cappi}, M. and {Malaguti}, P. and {Palumbo}, G.~G.~C. and {Risaliti}, G. and {Salvati}, M. and {Severgnini}, P.},
        title = "{An XMM-Newton hard X-ray survey of ultraluminous infrared galaxies}",
      journal = {\mnras},
         year = 2003,
        month = aug,
       volume = {343},
       number = {4},
        pages = {1181-1194},
          doi = {10.1046/j.1365-8711.2003.06744.x},
archivePrefix = {arXiv},
       eprint = {astro-ph/0304529},
 primaryClass = {astro-ph},
       adsurl = {https://ui.adsabs.harvard.edu/abs/2003MNRAS.343.1181F}
}

@ARTICLE{Vardoulaki_2015,
       author = {{Vardoulaki}, E. and {Charmandaris}, V. and {Murphy}, E.~J. and {Diaz-Santos}, T. and {Armus}, L. and {Evans}, A.~S. and {Mazzarella}, J.~M. and {Privon}, G.~C. and {Stierwalt}, S. and {Barcos-Mu{\~n}oz}, L.},
        title = "{Radio continuum properties of luminous infrared galaxies. Identifying the presence of an AGN in the radio}",
      journal = {\aap},
         year = 2015,
        month = feb,
       volume = {574},
          eid = {A4},
        pages = {A4},
          doi = {10.1051/0004-6361/201424125},
archivePrefix = {arXiv},
       eprint = {1408.4177},
 primaryClass = {astro-ph.GA},
       adsurl = {https://ui.adsabs.harvard.edu/abs/2015A&A...574A...4V}
}

@ARTICLE{Song_2022,
       author = {{Song}, Y. and {Linden}, S.~T. and {Evans}, A.~S. and {Barcos-Mu{\~n}oz}, L. and {Murphy}, E.~J. and {Momjian}, E. and {D{\'\i}az-Santos}, T. and {Larson}, K.~L. and {Privon}, G.~C. and {Huang}, X. and {Armus}, L. and {Mazzarella}, J.~M. and {U}, V. and {Inami}, H. and {Charmandaris}, V. and {Ricci}, C. and {Emig}, K.~L. and {McKinney}, J. and {Yoon}, I. and {Kunneriath}, D. and {Lai}, T.~S. -Y. and {Rodas-Quito}, E.~E. and {Saravia}, A. and {Gao}, T. and {Meynardie}, W. and {Sanders}, D.~B.},
        title = "{Characterizing Compact 15-33 GHz Radio Continuum Sources in Local U/LIRGs}",
      journal = {\apj},
         year = 2022,
        month = nov,
       volume = {940},
       number = {1},
          eid = {52},
        pages = {52},
          doi = {10.3847/1538-4357/ac923b},
archivePrefix = {arXiv},
       eprint = {2209.04002},
 primaryClass = {astro-ph.GA},
       adsurl = {https://ui.adsabs.harvard.edu/abs/2022ApJ...940...52S}
}

@ARTICLE{2020NumPy-Array,
  author  = {Harris, Charles R. and Millman, K. Jarrod and
            van der Walt, Stéfan J and Gommers, Ralf and
            Virtanen, Pauli and Cournapeau, David and
            Wieser, Eric and Taylor, Julian and Berg, Sebastian and
            Smith, Nathaniel J. and Kern, Robert and Picus, Matti and
            Hoyer, Stephan and van Kerkwijk, Marten H. and
            Brett, Matthew and Haldane, Allan and
            Fernández del Río, Jaime and Wiebe, Mark and
            Peterson, Pearu and Gérard-Marchant, Pierre and
            Sheppard, Kevin and Reddy, Tyler and Weckesser, Warren and
            Abbasi, Hameer and Gohlke, Christoph and
            Oliphant, Travis E.},
  title   = {Array programming with {NumPy}},
  journal = {Nature},
  year    = {2020},
  volume  = {585},
  pages   = {357–362},
  doi     = {10.1038/s41586-020-2649-2}
}

@ARTICLE{Matplotlib,
  author={Hunter, John D.},
  journal={Computing in Science \& Engineering}, 
  title={Matplotlib: A 2D Graphics Environment}, 
  year={2007},
  volume={9},
  number={3},
  pages={90-95},
  doi={10.1109/MCSE.2007.55}}

@article{Imanishi2009,
  author = {Imanishi, Masatoshi},
  title = {Buried AGNs in a complete sample of nearby ultraluminous infrared galaxies},
  journal = {Astronomical Society of the Pacific Conference Series},
  year = {2009},
  volume = {418},
  pages = {217--+},
  note = {ASPC “The AGN Unit”},
  doi = {?}
}

@article{Pfeifle2019a,
  author = {Pfeifle, Ryan W. and Secrest, Nathan J. and Satyapal, Shobita and Ellison, Sara L. and Rothberg, Barry and Blecha, Laura and Canalizo, Gabriela and Gliozzi, Mario},
  title = {Buried Black Hole Growth in IR-selected Mergers: New Results from Chandra},
  journal = {The Astrophysical Journal},
  year = {2019},
  volume = {875},
  number = {2},
  pages = {117},
  doi = {10.3847/1538-4357/ab0a93},
  archivePrefix = {arXiv},
  eprint = {1904.10955}
}

@article{Pfeifle2019b,
  author = {Pfeifle, Ryan W. and Secrest, Nathan J. and Satyapal, Shobita and Ellison, Sara L. and Rothberg, Barry and Blecha, Laura and Canalizo, Gabriela and Gliozzi, Mario},
  title = {A Triple AGN in a Mid-infrared-selected Late-stage Galaxy Merger},
  journal = {The Astrophysical Journal},
  year = {2019},
  volume = {883},
  number = {2},
  pages = {167},
  doi = {10.3847/1538-4357/ab3e6b},
  archivePrefix = {arXiv},
  eprint = {1908.01732}
}

@article{Efstathiou2022,
  author = {Efstathiou, Andreas and Farrah, Duncan and Afonso, J. and Clements, D. L. and González-Alfonso, E. and Lacy, M. and Oliver, S. and Papadopoulou Lesta, V. and Pearson, C. and Rigopoulou, D. and Rowan-Robinson, M. and Spoon, H. W. W. and Verma, A. and Wang, L.},
  title = {A new look at local ultraluminous infrared galaxies: the atlas and radiative transfer models of their complex physics},
  journal = {Monthly Notices of the Royal Astronomical Society},
  year = {2022},
  volume = {513},
  number = {4},
  pages = {4770–4790},
  doi = {10.1093/mnras/stac1555}
}

@article{Varnava2025,
  author = {Varnava, Charalambia and Efstathiou, Andreas and Farrah, Duncan and Rigopoulou, Dimitra},
  title = {Constraints on the active galactic nucleus and starburst activity of local ultraluminous infrared galaxies from a broad range of torus models},
  journal = {Monthly Notices of the Royal Astronomical Society},
  year = {2025},
  volume = {538},
  number = {1},
  pages = {426–442},
  doi = {10.1093/mnras/staf319}
}

@article{Farrah2022,
  author = {Farrah, Duncan and Efstathiou, Andreas and Afonso, J. and Bernard-Salas, J. and Cairns, J. and Clements, D. L. and Croker, K. and Hatziminaoglou, E. and Joyce, M. and Lacy, M. and Lebouteiller, V. and Lieblich, A. and Rowan-Robinson, M. and Runburg, J. and Spoon, H. W. W. and Verma, A. and Wang, L.},
  title = {Stellar and black hole assembly in z < 0.3 infrared-luminous mergers: intermittent starbursts vs. super-Eddington accretion},
  journal = {Monthly Notices of the Royal Astronomical Society},
  year = {2022},
  volume = {513},
  number = {4},
  pages = {4770–4790},
  doi = {10.1093/mnras/stac1555}
}
\bibliographystyle{aasjournalv7}



\end{document}